\documentclass[aps,prd,superscriptaddress,showpacs,nofootinbib]{revtex4}
\usepackage{graphicx,epsfig}
\usepackage[T1]{fontenc}
\usepackage{ae,aecompl}
\newcommand{\be}{\begin{equation}}\newcommand{\ee}{\end{equation}}
\newcommand{\ba}{\begin{eqnarray}}\newcommand{\ea}{\end{eqnarray}}
\newcommand{\la}{\langle}
\newcommand{\ra}{\rangle}
\newcommand{\Mn}{M_N} \newcommand{\di}{ {\rm d} }
\newcommand{\fracS}[2]{{\textstyle\frac{#1}{#2}}}
\newcommand{\btau}{   {{\mbox{\boldmath$\tau$} }}}
\newcommand{\bgam}{   {{\mbox{\boldmath$\gamma$}}}}
\newcommand{\bDelta}{ {{\mbox{\boldmath$\Delta$}}}}
\newcommand{\bnabla}{ {{\mbox{\boldmath$\nabla$}}}}
\newcommand{\fslash}[1] {{\not\! #1\,}}
\newcommand{\binomial}[2]{\left(\matrix{#1\cr #2}\right)}

\newcommand{\singlesum}[1]
        {\!\!\sum_{\renewcommand{\arraystretch}{0.3} \begin{array}{l}
    \scriptscriptstyle {#1} \\  \scriptscriptstyle {\phantom{n,\rm occ}}
    \end{array}}  \!\!}
\newcommand{\doublesum}[2]
    {\!\sum_{\renewcommand{\arraystretch}{0.3} \begin{array}{l}
    \scriptscriptstyle {#1} \\ \scriptscriptstyle {#2} \end{array}}}
\newcommand{\doublesumUp}[3]
    {\!\!\sum_{\renewcommand{\arraystretch}{0.3} \begin{array}{c}
    \scriptscriptstyle {#1} \\  \scriptscriptstyle {#2} \end{array}}
        ^{\scriptscriptstyle{#3}} \!\!}
\newcommand{\tripplesum}[3]
        {\!\!\sum_{\renewcommand{\arraystretch}{0.3} \begin{array}{c}
    \scriptscriptstyle {#1} \\ \scriptscriptstyle {#2} \\
    \scriptscriptstyle {#3} \end{array}}}
\newcommand{\doublearg}[2]
    {{  {\renewcommand{\arraystretch}{0.6} \begin{array}{c}
    \mbox{\scriptsize ${#1}$} \\ \mbox{\scriptsize ${#2}$}\end{array}}
    \!\!}}
\begin{document}
\newcommand*{\Bochum}{Institut f\"ur Theoretische Physik II,
Ruhr-Universit\"at Bochum, D-44780 Bochum, Germany}\affiliation{\Bochum}
\newcommand*{\Liege}{Universit\'e de Li\`ege au Sart Tilman,
B-4000 Li\`ege 1, Belgium}\affiliation{\Liege}
\newcommand*{\StPetersburg}{Petersburg Nuclear Physics Institute, Gatchina,
St.~Petersburg 188350, Russia}\affiliation{\StPetersburg}
\newcommand*{\Porto}{Faculdade de Engenharia da Universidade do Porto,
P-4000 Porto, Portugal}\affiliation{\Porto}
\newcommand*{\Coimbra}{Centro de F\'isica Computacional, Universidade de 
Coimbra, P-3000 Coimbra, Portugal}
\title{ \boldmath
        The generalized parton distribution function $(E^u+E^d)(x,\xi,t)$
        \\
        of the nucleon in the chiral quark soliton model}
\author{J.~Ossmann}\affiliation{\Bochum}
\author{M.~V.~Polyakov}\affiliation{\Liege}\affiliation{\StPetersburg}
\author{P.~Schweitzer}\affiliation{\Bochum}
\author{D.~Urbano}\affiliation{\Porto}\affiliation{\Coimbra}
\author{K.~Goeke}\affiliation{\Bochum}
\date{January 2005}
\begin{abstract}
The unpolarized spin-flip isoscalar generalized parton distribution function
$(E^u+E^d)(x,\xi,t)$ is studied in the large-$N_c$ limit at a low
normalization point in the framework of the chiral quark-soliton model.
This is the first study of generalized parton distribution functions in this
model, which appear only at the subleading order in the large-$N_c$ limit.
Particular emphasis is put therefore on the demonstration of the
theoretical consistency of the approach. The forward limit of
$(E^u+E^d)(x,\xi,t)$ of which only the first moment -- the anomalous
isoscalar magnetic moment of the nucleon -- is known phenomenologically,
is computed numerically.
Observables sensitive to $(E^u+E^d)(x,\xi,t)$ are discussed.
\end{abstract}
\pacs{13.60.Hb, 12.38.Lg, 12.39.Ki, 14.20.Dh}
\maketitle

\vspace{-10cm}\begin{flushright}
{\tt preprint RUB-TPII-6/04}
\end{flushright}\vspace{9cm}

  \section{Introduction}
  \label{Sec:1-introduction}

The study of generalized parton distribution functions (GPDs) \cite{first-GPD}
(see \cite{Ji:1998pc,Radyushkin:2000uy,Goeke:2001tz,Diehl:2003ny} for reviews)
promises numerous new insights in the internal structure of the nucleon
\cite{Ji:1996ek,interpret,Polyakov:2002yz}. GPDs can be accessed in a variety
of hard exclusive processes \cite{Vanderhaeghen:2002ax} on which recently
first data became available \cite{experiments}.
Current efforts \cite{efforts} to understand and interpret the data rely,
at the early stage of art, on modeling Ans\"atze for GPDs -- which have to
comply with the severe general constraints imposed by the polynomiality and
positivity properties of GPDs \cite{Polyakov:2002wz,positivity}.
Model calculations \cite{Ji:1997gm,Petrov:1998kf,Penttinen:1999th,Schweitzer:2002nm,Mukherjee:2002pq,Boffi:2002yy,Radyushkin:2004mt,Vanderhaeghen:Festschrift}
can provide important guidelines for such Ans\"atze  -- in particular in the
case of those GPDs for which not even the forward limit is known from deeply
inelastic scattering experiments.
In this work we study the unpolarized GPD $E^a(x,\xi,t)$ in the chiral quark
soliton model ($\chi$QSM) \cite{Diakonov:yh,Diakonov:1987ty}.

The model describes the nucleon
in a field theoretic framework in the limit of a large number of colours $N_c$
as a chiral soliton of a static background pion field.
Numerous nucleonic properties -- among others form factors
\cite{Christov:1995hr,Christov:1995vm} as well as quark and antiquark
distribution functions
\cite{Diakonov:1996sr,Pobylitsa:1996rs,Diakonov:1997vc,Weiss:1997rt,Pobylitsa:1998tk,Diakonov:1998ze,Wakamatsu:1998rx,Goeke:2000wv,Schweitzer:2001sr}
-- have been described in this model without adjustable
parameters, typically to within an accuracy of $(10-30)\%$.
The field-theoretical character is a crucial feature of the $\chi$QSM which
is responsible for the wide range of applicability and which guarantees the
theoretical consistency of the approach.
In Refs.~\cite{Petrov:1998kf,Penttinen:1999th,Schweitzer:2002nm} it was
demonstrated that the $\chi$QSM consistently describes those GPDs, which 
are of leading order in the large $N_c$-limit.

In this work we extend the formalism of Ref.~\cite{Petrov:1998kf} to the
description of unpolarized GPDs which appear only at subleading order in the
large-$N_c$ expansion. We pay particular attention to the demonstration of the
consistency of the approach. After checking explicitly that the model
expressions for GPDs satisfy polynomiality, relevant sum rules, etc.\  we
focus on the flavour combination of $E^a(x,\xi,t)$ subleading in large-$N_c$,
namely the flavour singlet, which -- being related to the spin sum rule
of the nucleon -- is of particular phenomenological interest.
The corresponding  flavour structure $(E^u-E^d)(x,\xi,t)$, which is leading 
in $N_c$, was already studied in Ref.~\cite{Petrov:1998kf} ({\sl cf.} 
\cite{Goeke:2001tz}).
An interesting issue we shall finally address is, which hard exclusive
reactions are particularly sensitive to the GPD $(E^u+E^d)(x,\xi,t)$.
We observe that target single transverse spin asymmetries in two pion
production are most promising in this respect.

This note is organized as follows. After a general discussion of the
properties of GPDs in Sec.~\ref{Sec:2-GPDs-general} and a short introduction
to the $\chi$QSM in Sec.~\ref{Sec:3-model}, we review in
Sec.~\ref{Sec:4-GPDs-in-the-model} the formalism of the $\chi$QSM for GPDs
of leading order in $N_c$.
In Sec.~\ref{Sec:5-GPDs-in-NLO} we generalize the approach to the case of
GPDs subleading in $N_c$, and check its consistency.
In Sec.~\ref{Sec:6-Eu+Ed-forward-lim} we apply the approach to the numerical
calculation of the forward limit of $(E^u+E^d)(x,\xi,t)$ and discuss
phenomenological implications from our result in Sec.~\ref{Sec:7-asymmetries}.
Finally, Sec.~\ref{Sec:8-conclusions} contains a summary and conclusions.
Technical details on the calculations can be found in the Appendices.

\newpage
  \section{The unpolarized generalized parton distribution functions}
  \label{Sec:2-GPDs-general}

The unpolarized quark GPDs in the proton are defined as
\ba\label{Def:1} &&  \hspace{-1cm}
    \int\!\frac{\di\lambda}{2\pi}\,e^{i\lambda x}\la {\bf P'},s'|
    \bar{\psi}_q(-\fracS{\lambda n}{2}) \fslash{n}
    [-\fracS{\lambda n}{2},\fracS{\lambda n}{2}]
    \psi_q(\fracS{\lambda n}{2})|{\bf P},s\ra \nonumber\\
&&  =  H^q(x,\xi,t)\;\bar{U}({\bf P'},s')\fslash{n}U({\bf P},s)
    \nonumber\\
&&  +  E^q(x,\xi,t)\;
    \bar{U}({\bf P'},s')\,\frac{i\sigma^{\mu\nu}n_\mu\Delta_\nu}{2\Mn}\,
    U({\bf P},s) , \ea
where $[z_1,z_2]$ denotes the gauge-link,
and the renormalization scale dependence is not indicated for brevity.
The light-like vector $n^\mu$ satisfies $n(P'+P) = 2$.
The skewedness parameter $\xi$, the four-momentum transfer
$\Delta^\mu$ and the Mandelstam variable $t$ are defined as
$\Delta^\mu = (P'-P)^\mu$, $n\Delta = -2\xi$, $t = \Delta^2$.
The antiquark distributions are given by $H^{\bar q}(x,\xi,t)=-H^q(-x,\xi,t)$
and similarly for $E^{\bar q}(x,\xi,t)$.
The unpolarized GPDs are normalized
to the corresponding elastic (Dirac- and Pauli-) form factors
\ba\label{Eq:GPD-form-factor}
    \int\limits_{-1}^1\!\di x\; H^q(x,\xi,t) = F_1^q(t)\;,\;\;\;
    \int\limits_{-1}^1\!\di x\; E^q(x,\xi,t) = F_2^q(t)\;.
    \label{Eq:form-factors}\ea
The relations (\ref{Eq:GPD-form-factor}) are special cases of the
polynomiality property \cite{Ji:1998pc} which follows from hermiticity,
parity, time reversal and Lorentz invariance, and implies that the
$N^{\rm th}$ Mellin moment of an unpolarized GPDs is a polynomial in
even powers of $\xi$ of degree less than or equal to $N$
\ba
    \int\limits_{-1}^1\!\!\di x\:x^{N-1}\,H^q(x,\xi,0)
    = h^{q\,(N)}_0 + h^{q\,(N)}_2 \xi^2 + \dots +
        \cases{h^{q\,(N)}_N    \xi^N    \!\!\!\!\! & for $N$ even\cr
           h^{q\,(N)}_{N-1}\xi^{N-1}\!\!\!\!\! & for $N$ odd,} &&
    \label{Def:polynom-Hq}\\
    \int\limits_{-1}^1\!\!\di x\:x^{N-1}\,E^q(x,\xi,0)
    = e^{q\,(N)}_0 + e^{q\,(N)}_2 \xi^2 + \dots +
        \cases{e^{q\,(N)}_N    \xi^N    \!\!\! & for $N$ even\cr
           e^{q\,(N)}_{N-1}\xi^{N-1}\!\!\! & for $N$ odd.}&&
    \label{Def:polynom-Eq} \ea
As a consequence of the spin $\frac12$ nature of the nucleon the coefficients
in front of the highest power in $\xi$ for even moments $N$ are related to
each other by
\be
    h^{q\,(N)}_N=-\,e^{q\,(N)}_N=\int\limits_{-1}^1\di z\; z^{N-1}D^q(z)
        \;\;,\label{Eq:relation-h-e}
\ee
and arise from the so-called $D$-term $D^q(z)$ with $z=x/\xi$ which has
finite support only for $|x|<|\xi|$ \cite{D-term}. The $D$-term governs
the asymptotics of unpolarized GPDs in the limit of a large renormalization
scale \cite{Goeke:2001tz} and is related to the distribution of the pressure
and shear forces acting on the partons in the nucleon \cite{Polyakov:2002yz}.

In the large-$N_c$ limit different flavour combinations of
GPDs and of the D-term exhibit the behaviour \cite{Goeke:2001tz}
\ba
    (H^u+H^d)(x,\xi,t) = N_c^2 f(N_cx,N_c\xi,t) \;, &&
    (H^u-H^d)(x,\xi,t) = N_c\, f(N_cx,N_c\xi,t) \;, \nonumber\\
    (E^u+E^d)(x,\xi,t) = N_c^2 f(N_cx,N_c\xi,t) \;, &&
    (E^u-E^d)(x,\xi,t) = N_c^3 f(N_cx,N_c\xi,t) \;, \nonumber\\
    (D^u+D^d)(z)\phantom{,\xi,t}= N_c^2 f(z)\phantom{N_c,N_c\xi,t} \;, &&
    (D^u-D^d)(z)\phantom{,\xi,t} = N_c\, f(z)\;\;.
    \label{Eq:largeNc}\ea
The functions $f(u,v,t)$ and $f(z)$ (which are of order ${\cal O}(N_c^0)$
and which we do not distinguish for notational simplicity) are stable in
the large-$N_c$ limit for fixed values of $u,\,v,\,z,\,t={\cal O}(N_c^0)$.
$E^q(x,\xi,t)$ is systematically enhanced by one order in $N_c$ with
respect to $H^q(x,\xi,t)$ due to the nucleon mass $\Mn={\cal O}(N_c)$
appearing in the denominator on the right hand side of Eq.~(\ref{Def:1}).
As a consequence the flavour combination of $H^q(x,\xi,t)$ leading in $N_c$,
namely $(H^u+H^d)(x,\xi,t)$, and the flavour combination of $E^q(x,\xi,t)$
subleading in $N_c$, namely $(E^u+E^d)(x,\xi,t)$, have the same order
in $N_c$. This is natural from the point of view of
the spin sum rule \cite{Ji:1996ek}
    \be\label{Eq:spin-sum-rule}
    \int_{-1}^1\di x\;\,x \!
    \sum\limits_{q=u,d,\dots} (H^q+E^q)(x,\xi,t)
    = 2 J^Q\;, \ee
where $2J^Q$ is the fraction of the nucleon spin due to (spin and
orbital angular momentum of) quarks. The flavour singlets of
$E^q(x,\xi,t)$ and $H^q(x,\xi,t)$ enter the left-hand-side of
(\ref{Eq:spin-sum-rule}) and contribute on equal footing in
the large-$N_c$ limit to $J^Q={\cal O}(N_c^0)$.

In the forward limit of $H^q(x,\xi,t)$ we recover the unpolarized
parton distribution function $f_1^q(x)$
    \be\label{Eq:forward-limit-H}
    \lim\limits_\doublearg{\xi\to 0}{t\to 0}
    H^q(x,\xi,t) = f_1^q(x)\;. \ee
The GPD $E^q(x,\xi,t)$ has also a well defined forward limit which,
however, is not accessible in deeply inelastic lepton nucleon scattering.
Phenomenologically only the first moment of $E^q(x,0,0)$ is known which
is given by the anomalous magnetic moment $\kappa^q \equiv F_2^q(0)$ with
    \be\label{Eq:kappa}
    \int\limits_0^1\di x\,(E^u-E^{\bar u})(x,0,0)= \kappa_u = 1.673 \, , \;\;\;
    \int\limits_0^1\di x\,(E^d-E^{\bar d})(x,0,0)= \kappa_d = -2.033 \, . \ee

  \section{The chiral quark-soliton model \boldmath ($\chi$QSM)}
  \label{Sec:3-model}

The effective chiral relativistic field theory underlying the
$\chi$QSM is given by the partition function
\cite{Diakonov:tw,Diakonov:1985eg} \be
    Z_{\rm eff} = \int\!{\cal D}\psi\,{\cal D}\bar{\psi}\,{\cal D}U\;
    \exp\Biggl(i\int\!\di^4x\;\bar{\psi}\,
    (i\fslash{\partial}-M\,U^{\gamma_5})\psi\Biggr) \,,
    \label{Eq:eff-theory} \ee
where $\psi$ and $U=\exp(i\tau^a\pi^a)$ are the $SU(2)$ chiral quark and pion
fields with $U^{\gamma_5}\equiv\exp(i\gamma_5\tau^a\pi^a)$, and $M$ is the
dynamical quark mass. 
The effective theory (\ref{Eq:eff-theory}) was derived from the instanton 
model of the QCD vacuum \cite{Diakonov:1985eg,Diakonov:1983hh}. An important
small parameter in this derivation is the instanton packing fraction
$\rho_{\rm av}/R_{\rm av}\sim \frac13$ of the dilute instanton medium,
where $\rho_{\rm av}$ and $R_{\rm av}$ are respectively the average size and 
separation of instantons. The effective theory (\ref{Eq:eff-theory}) contains 
the Weinberg-Gasser-Leutwyler Lagrangian and the Wess-Zumino term with correct
coefficients and is valid for momenta below a scale set by the inverse of 
the average instanton size $\rho_{\rm av}^{-1}\approx 600\,{\rm MeV}$.
At this scale the dynamical quark mass $M$, which in general is momentum
dependent, drops to zero. In numerical calculations it is often convenient to
consider constant $M$ and to apply an appropriate regularization method with
a UV-cutoff of the order of magnitude of $\rho_{\rm av}^{-1}$.
In calculations of (some) GPDs, however, it is important to consider momentum
dependent $M$ \cite{Petrov:1998kf}.

In the large-$N_c$ limit, which allows to solve the functional integral over
pion field configurations in the saddle-point approximation, the effective
theory in Eq.~(\ref{Eq:eff-theory}) describes the nucleon as a classical
soliton of the pion field \cite{Diakonov:yh} providing a practical
realization of the large-$N_c$ picture of the nucleon \cite{Witten:tx}.
Quarks are described by one-particle wave functions which are the solutions
of the Dirac equation in the background of the static pion field
\be
    \hat{H}_{\rm eff}|n\ra = E_n |n\ra \;\;,\;\;\;\;
    \hat{H}_{\rm eff} = -i\gamma^0\gamma^k\partial_k+\gamma^0MU^{\gamma_5}
    \;.\label{Eq:Hamiltonian}\ee
The spectrum of the effective Hamiltonian (\ref{Eq:Hamiltonian}) consists
of an upper and a lower Dirac continuum, which are distorted by the
background field compared to the continua of the free Hamiltonian
$\hat{H}_0 = -i\gamma^0\gamma^k\partial_k+\gamma^0 M$,
and of a discrete bound state level of energy $E_{\rm lev}$.
By occupying the discrete level and the states of lower continuum each
by $N_c$ quarks in an anti-symmetric colour state, one obtains a state
with unity baryon number called soliton.
The minimization of the soliton energy $E_{\rm sol}$ with respect to
variations of the chiral field $U$ yields the self-consistent pion field
$U_c$, which for symmetry reasons has the ``hedgehog'' structure
$U_c({\bf x})=\exp[i{\bf e_r}\btau P(r)]$ where $P(r)$ is the soliton
profile with $r=|{\bf x}|$ and ${\bf e_r}={\bf x}/r$. The mass of the
nucleon is given by
\ba
&&  \Mn = E_{\rm sol}[U_c] = \min\limits_U E_{\rm sol}[U]\,,\;\;\;
    E_{\rm sol}[U] =
    N_c \biggl(E_{\rm lev}+\singlesum{E_n<0}(E_n-E_{n_0})\biggr)_{\rm reg}
    \;.\label{Eq:sol-energy}\ea
The soliton energy is (for constant $M$) logarithmically UV-divergent
and has to be regularized -- as indicated in Eq.~(\ref{Eq:sol-energy})
and described in Ref.~\cite{Christov:1995vm}.

In order to include $1/N_c$ corrections one has to consider quantum
fluctuations around the saddle point solution. Hereby the (translational
and rotational) zero modes of the soliton solution 
are the only taken into account.
In particular one considers time-dependent rotations of the hedgehog field,
$U_c({\bf x})\to R(t)U_c({\bf x})R^\dag(t)$, where the collective coordinate
$R(t)$ is a rotation matrix in SU(2)-flavour space. The path integration over
the collective coordinates can be solved by expanding in powers of the
collective angular velocity $\Omega\equiv -iR^\dag \partial_t R$ which
corresponds to an $1/N_c$ expansion. The latter is justified, since the soliton
moment of inertia,
\be\label{Eq:mom-inertia}
    I = \frac{N_c}{6}\,\doublesum{n,\,\rm occ}{j,\,\rm non}
    \frac{\la n|\tau^a|j\ra\la j|\tau^a|n\ra}{E_j-E_n}, \ee
is large, namely $I={\cal O}(N_c)$, and the soliton rotation is therefore slow.
(In Eq.~(\ref{Eq:mom-inertia}) one has to sum over occupied ("occ") states $n$,
i.e.\  over states with $E_n\le E_{\rm lev}$, and over non-occupied ("non")
states $j$, i.e.\  over states with $E_j  > E_{\rm lev}$.)
This procedure -- which is referred to as quantization of zero modes --
assigns to the soliton a definite momentum and spin-isospin quantum numbers
\cite{Diakonov:yh}.

The effective theory (\ref{Eq:eff-theory}) allows to derive by means of path
integral methods unambiguous model expressions for nucleon matrix elements
of QCD quark bilinear operators sandwiched in nucleon states
$\la N'|\bar{\psi}(z_1)[z_1,z_2]\Gamma\psi(z_2)|N\ra$, where
$\Gamma$ is some matrix in Dirac- and flavour space.
For local observables such as, e.g., form factors the gauge link reduces to
a unity matrix in colour space. When evaluating non-local operators in the
model -- as they appear in GPDs -- it is crucial that the effects of gluonic
degrees of freedom, which are intrinsic in the gauge link, appear strongly
suppressed with respect to quark degrees of freedom. This is true for twist-2
\cite{Diakonov:1995qy}
(and certain twist-3 operators \cite{Balla:1997hf}) and guarantees the colour gauge
invariance of the model calculation.

If in QCD $\la N'|\bar{\psi}(z_1)[z_1,z_2]\Gamma\psi(z_2)|N\ra$
is scale dependent then the model result corresponds to low a scale of
$\rho^{-1}\approx 600\,{\rm MeV}$. In this way static nucleonic observables
\cite{Christov:1995hr,Christov:1995vm}, twist-2 quark and anti-quark
distribution functions
\cite{Diakonov:1996sr,Pobylitsa:1996rs,Diakonov:1997vc,Weiss:1997rt,Pobylitsa:1998tk,Diakonov:1998ze,Wakamatsu:1998rx,Goeke:2000wv,Schweitzer:2001sr}
have been computed in the $\chi$QSM and found to agree to within $(10-30)\%$
with experimental data or phenomenological parameterizations.
In \cite{Petrov:1998kf,Penttinen:1999th} the approach was generalized to
describe GPDs. The results of the $\chi$QSM respect all general counting
rules of the large-$N_c$ phenomenology.

  \section{GPDs in leading order of large \boldmath $N_c$}
  \label{Sec:4-GPDs-in-the-model}

It is convenient to treat the cases of flavour singlet and non-singlet
quantities in the model separately for symmetry reasons.
Let us introduce the notation
\ba
    {\cal M}^{(I=0)}_{s^\prime s}
    &\equiv&
  \int\!\frac{\di\lambda}{2\pi}\,e^{i\lambda x}\la{\bf P^\prime},s^\prime|
  \bar\psi(-\fracS{\lambda n}{2})\,\fslash{n}\,\psi(\fracS{\lambda n}{2})
  |{\bf P},s\ra
    \label{Def:MI-0}\\
    {\cal M}^{(I=1)}_{s^\prime s}
    &\equiv&
  \int\!\frac{\di\lambda}{2\pi}\,e^{i\lambda x}\la {\bf P^\prime },s^\prime |
  \bar\psi(-\fracS{\lambda n}{2})\,\tau^3\fslash{n}\,\psi(\fracS{\lambda n}{2})
  |{\bf P},s\ra  \;.  \label{Def:MI-1}\ea
The ${\cal M}^{(I)}_{s^\prime s}$ are $2\times 2$-matrices in spin indices
which have the following behaviour in the large-$N_c$ limit
\ba
    {\rm tr}\,\{{\cal M}^{(I=0)}\}         = {\cal O}(N_c^2)\;, &&
    {\rm tr}\,\{\sigma^m{\cal M}^{(I=0)}\} = {\cal O}(N_c  )\;,\nonumber\\
    {\rm tr}\,\{{\cal M}^{(I=1)}\}         = {\cal O}(N_c  )\;, &&
    {\rm tr}\,\{\sigma^m{\cal M}^{(I=1)}\} = {\cal O}(N_c^2)\;.
    \label{Eq:Nc-counting-0} \ea
These relations do not follow from the dynamics of the model, but are group
theoretical consequences of the spin-flavour structure of the (hedgehog)
soliton field. In this sense the relations (\ref{Eq:Nc-counting-0}) are
model-independent large-$N_c$ results of QCD \cite{Ref:Nc-counting} which
are consequently respected in the $\chi$QSM.

In order to evaluate the bi(nucleon-)spinor expression on the right-hand-side
of Eqs.~(\ref{Def:1}) one has to consider that in the large $N_c$ limit the
nucleon is heavy, $\Mn={\cal O}(N_c)$, and the kinematics becomes
non-relativistic. In particular we have for the components of the momentum
transfer $\Delta^i = {\cal O}(N_c^0)$ and $\Delta^0 = {\cal O}(N_c^{-1})$.
Thus the hierarchy holds $\Mn\gg |\Delta^i| \gg |\Delta^0|$, while
$t=-\bDelta^2={\cal O}(N_c^0)$ and $\xi=-\Delta^3/(2\Mn)={\cal O}(N_c^{-1})$.
Here we have chosen $n^\mu= (1,\,0,\,0,-1)/\Mn$.
Evaluating consequently the right-hand-side of Eq.~(\ref{Def:1}) in this
large-$N_c$ kinematics yields for the flavour-singlet and non-singlet case,
respectively,
\ba
    {\cal M}^{(I=0)}_{s^\prime s}
    &=& 2\,\delta_{ss^\prime }\;(H^u+H^d)(x,\xi,t)
         -  \frac{i\epsilon^{3kl}\Delta^k}{\Mn}(\sigma^l)_{s^\prime s}
        \biggl[(H^u+H^d)(x,\xi,t)+(E^u+E^d)(x,\xi,t)\biggr] \;,
    \label{Def:4b}\\
    {\cal M}^{(I=1)}_{s^\prime s}
    &=& 2\,\delta_{ss^\prime }\;\biggl[(H^u-H^d)(x,\xi,t)
         +  \frac{t}{4\Mn^2}(E^u-E^d)(x,\xi,t)\biggr]
         -  \frac{i\epsilon^{3kl}\Delta^k}{\Mn}(\sigma^l)_{s^\prime s}
            (E^u-E^d)(x,\xi,t) \;.\nonumber\\
    \label{Def:4c}\ea
From Eqs.~(\ref{Def:4b},~\ref{Def:4c}) simple relations follow
for the leading (in the large-$N_c$ counting) GPDs
\ba
&&\label{Hu+Hd-00}
  (H^u+H^d)(x,\xi,t)=\frac{1}{4}\,{\rm tr}\{{\cal M}^{(I=0)}\}\;, \\
&&\label{Eu-Ed-00}
  (E^u-E^d)(x,\xi,t)=\frac{i\Mn\epsilon^{3bm}\Delta^b\!}{2\,\bDelta^2_\perp}\,
                     {\rm tr}\,\{\sigma^m {\cal M}^{(I=1)}\}\;,
\ea
where ``${\rm tr}$'' denotes the trace over spin indices and
$\bDelta^2_\perp=\bDelta^2-(\Delta^3)^2=-t-4\Mn^2\xi^2$. The subleading
structures, however, appear only in combination with the leading ones
\ba
&& \label{Eu+Ed-00} (H^u+H^d)(x,\xi,t)+(E^u+E^d)(x,\xi,t)
   = \frac{i\Mn\epsilon^{3bm}\Delta^b\!}{2\,\bDelta^2_\perp}\,
     {\rm tr}\{\sigma^m{\cal M}^{(I=0)}\}\;,\\
&& \label{Hu-Hd-00} (H^u-H^d)(x,\xi,t)+\frac{t}{4\Mn^2}\,(E^u-E^d)(x,\xi,t)
   = \frac{1}{4}{\rm tr}\{{\cal M}^{(I=1)}\}\;.
\ea
The structures in the square brackets of Eqs.~(\ref{Def:4b},~\ref{Def:4c})
and on the left-hand-sides of Eqs.~(\ref{Eu+Ed-00},~\ref{Hu-Hd-00}) are
analogs of the relations between the electric and magnetic form factors
$G_E^q(t),\, G_M^q(t)$ and the Dirac- and Pauli-form factors
$F_1^q(t),\,F_2^q(t)$, which are given by
\be\label{Eq:form-factor-relations}
    G_E^q(t)=F_1^q(t)+F_2^q(t)\,,\;\;\;
    G_M^q(t)=F_1^q(t)+\frac{t}{4\Mn^2}\,F_2^q(t)\,.\ee

The model expressions for ${\cal M}^{(I=0,1)}$ in leading order of the
large-$N_c$ limit were derived in Ref.~\cite{Petrov:1998kf} and the
following results for $(H^u+H^d)(x,\xi,t)$ and $(E^u-E^d)$ were obtained
using Eqs.~(\ref{Def:4b},~\ref{Def:4c})
\ba
     (H^u+H^d)(x,\xi,t)
&=&     \Mn N_c\int\frac{\di z^0}{2\pi}\singlesum{n,\rm occ}e^{iz^0(x\Mn-E_n)}
        \nonumber\\
&&   \times
    \la n|\,(1+\gamma^0\gamma^3)\exp(-i\fracS{z^0}{2}\hat{p}^3)
    \exp(i\bDelta\hat{\bf X})\exp(-i\fracS{z^0}{2}\hat{p}^3)\,|n\ra
    \label{Hu+Hd-model}\;,\\
    \phantom{X}\nonumber\\
    (E^u-E^d)(x,\xi,t)
&=&     \frac{2i\Mn^2 N_c}{3(\bDelta^{\!\perp})^2}
        \int\frac{\di z^0}{2\pi} \singlesum{n,\rm occ} e^{iz^0(x\Mn - E_n)}
        \nonumber\\
&&  \times
    \la n|\,(1+\gamma^0\gamma^3)(\btau\times\bDelta)^3
    \exp(-i\fracS{z^0}{2}\hat{p}^3)\exp(i\bDelta\hat{\bf X})
    \exp(-i\fracS{z^0}{2}\hat{p}^3)\,|n\ra
    \label{Eu-Ed-model}\;.\ea
There are equivalent expressions with $\sum_{\rm occ}\to -\sum_{\rm non}$,
i.e.\  where the summation goes over non-occupied states $E_n > E_{\rm lev}$.
The possibility of computing in the model quantities in these two independent
ways is deeply related to the locality properties of the model
\cite{Diakonov:1996sr}.
Numerical results for (\ref{Hu+Hd-model},~\ref{Eu-Ed-model}) were presented
in Ref.~\cite{Petrov:1998kf}, and further discussed and reviewed in
Ref.~\cite{Goeke:2001tz}.

Let us emphasize that according to Eq.~(\ref{Eq:forward-limit-H}) in the
forward limit the right-hand-side of  (\ref{Hu+Hd-model}) reduces to the
model expression for $(f_1^u+f_1^d)(x)$ \cite{Petrov:1998kf}.
$(H^u+H^d)(x,\xi,t)$ and $(E^u-E^d)(x,\xi,t)$ are correctly normalized
to the respective form factors \cite{Petrov:1998kf}, {\sl cf.}
Eq.~(\ref{Eq:GPD-form-factor}), and they satisfy the polynomiality
conditions in Eqs.~(\ref{Def:polynom-Hq},~\ref{Def:polynom-Eq})
\cite{Schweitzer:2002nm}.

It is worthwhile mentioning that 
the coefficient of the highest power in $\xi$ of even Mellin moments of
$(E^u-E^d)(x,\xi,t)$ in leading order of large $N_c$ is zero in the $\chi$QSM.
Also this result is not a dynamical feature of the model,
but rather a group theoretical consequence of the soliton symmetries which
ensure the correct large-$N_c$ counting for the flavour-nonsinglet $D$-term.
In fact, if this coefficient were not zero, then $(E^u-E^d)={\cal O}(N_c^3)$
would imply $(D^u-D^d)={\cal O}(N_c^3)$ in conflict with the counting rule
which states $(D^u-D^d)={\cal O}(N_c)$, {\sl cf.} Eq.~(\ref{Eq:largeNc}).

  \section{GPDs in subleading order of large \boldmath $N_c$}
  \label{Sec:5-GPDs-in-NLO}

One has to consider $1/N_c$ (rotational) corrections to ${\cal M}^{(I=0,1)}$
in order to study the subleading flavour combinations $(H^u-H^d)(x,\xi,t)$
and $(E^u+E^d)(x,\xi,t)$. In order to simplify the notation let us introduce
\ba
\label{Def:E_M} && E_M(x,\xi,t)=(H^u+H^d)(x,\xi,t)+(E^u+E^d)(x,\xi,t)
        \sim{\cal O}(N_c^2)\;,\\
\label{Def:H_E} && H_E(x,\xi,t)=(H^u-H^d)(x,\xi,t)+(E^u-E^d)(x,\xi,t)
        \,\frac{t}{4\Mn^2}\sim{\cal O}(N_c)\;.\ea
Here the respective already known (see above) flavour combinations of
$H^q(x,\xi,t)$ and $E^q(x,\xi,t)$ leading in $N_c$ appear. Thus it is
sufficient to focus on the ``new'' objects $E_M(x,\xi,t)$ and $H_E(x,\xi,t)$.

The model expressions for $H_E(x,\xi,t)$ and $E_M(x,\xi,t)$ (for a proton)
read
\ba
    H_E(x,\xi,t) &=&
    -\frac{\Mn N_c}{12I}\;\int\!\frac{\di z^0}{2\pi}
    \biggl[\biggl\{
     \tripplesum{m,{\rm occ}}{j,{\rm all}}{m\neq j}e^{-iz^0E_m}
        -\tripplesum{m,{\rm all}}{j,{\rm occ}}{m\neq j}e^{-iz^0E_j}\biggr\}
     \frac{1}{E_m-E_j}
    +\frac{\partial\;\;}{\partial x\Mn}
     \tripplesum{m,{\rm occ}}{j,{\rm all}}{m\neq j}e^{-iz^0E_m}\biggr]
    e^{iz^0x\Mn}
     \nonumber\\
&&  \times\,\la m|\tau^a|j\ra\,
        \la j|\,\tau^a(1+\gamma^0\gamma^3)\,\exp(-iz^0\hat{p}^3/2)\,
    \exp(i\bDelta\hat{\bf X})\,\exp(-iz^0\hat{p}^3/2)\,|m\ra\;,
    \label{H_E-model}\\ &&\phantom{X}\nonumber\\
    E_M(x,\xi,t) &=&
         \frac{i\Mn^2 N_c}{2I}\;\int\!\frac{\di z^0}{2\pi}
    \biggl[\biggl\{
     \tripplesum{m,{\rm occ}}{j,{\rm all}}{m\neq j}e^{-iz^0E_m}
        -\tripplesum{m,{\rm all}}{j,{\rm occ}}{m\neq j}e^{-iz^0E_j}\biggr\}
     \frac{1}{E_m-E_j}
    +\frac{\partial\;\;}{\partial x\Mn}
     \tripplesum{m,{\rm occ}}{j,{\rm all}}{m\neq j}e^{-iz^0E_m}\biggr]
    e^{iz^0x\Mn}
    \nonumber\\
&&  \times\,\la m|\tau^b|j\ra\,
        \la j|\,(1+\gamma^0\gamma^3)\,\exp(-iz^0\hat{p}^3/2)\,
        \frac{\epsilon^{3ab} \Delta^a}{\bDelta_\perp^2}\;
    \exp(i\bDelta\hat{\bf X})\,\exp(-iz^0\hat{p}^3/2)\,|m\ra\;.
    \label{E_M-model}\ea
There are equivalent expressions with opposite sign where the summations
go over non-occupied states. Considering the generalizations due to the
off-forward kinematics \cite{Petrov:1998kf,Penttinen:1999th}, the derivations
of Eqs.~(\ref{H_E-model},~\ref{E_M-model}) closely follow the derivations of
the model expression for the flavour non-singlet unpolarized distribution
$(f_1^u-f_1^d)(x)$ \cite{Pobylitsa:1998tk} and the flavour-singlet helicity
$(g_1^u+g_1^d)(x)$ and transversity $(h_1^u+h_1^d)(x)$ distribution functions
\cite{Goeke:2000wv}.
In the following we check explicitly the theoretical consistency of the
expressions (\ref{H_E-model},~\ref{E_M-model}).

\subsection{Form factors and polynomiality}

Let us first verify the correct normalization of $H_E(x,\xi,t)$ and
$E_M(x,\xi,t)$. In order to integrate Eqs.~(\ref{H_E-model},~\ref{E_M-model})
over $x$ it is convenient to substitute $x\to y=x\Mn$ and to extend the
$y$-integration range $[-\Mn,\Mn]$ to $[-\infty,\infty]$ in the large-$N_c$
limit.
The derivative in $x$ drops out and after cancellations in the curly brackets
in Eqs.~(\ref{H_E-model},~\ref{E_M-model}) only summations over occupied
states $|m\ra$ and non-occupied states $|j\ra$ and vice versa remain, which
can be combined by exploring model symmetries, see App.~\ref{App:symmetries}.
We obtain
\ba
&& \int\limits_{-1}^1 \di x \, H_E(x,\xi,t) =
    -\;\frac{N_c}{6I}\doublesum{m,\rm occ}{j,\rm non}
    \frac{1}{E_m-E_j}\,\la m|\tau^a|j\ra\,
        \la j|\,\tau^a\,\exp(i\bDelta\hat{\bf X})\,|m\ra
    \equiv G_E^{(I=1)}(t) \;,\nonumber\\
&& \int\limits_{-1}^1 \di x \, E_M(x,\xi,t) =
    \frac{N_c\Mn}{2I\bDelta^2}\doublesum{m,\rm occ}{j,\rm non}
    \frac{i\epsilon^{abc}\Delta^a}{E_m-E_j}\,\la m|\tau^b|j\ra\,
        \la j|\,\gamma^0\gamma^c\exp(i\bDelta\hat{\bf X})\,|m\ra
    \equiv 3\,G_M^{(I=0)}(t)\;, \label{Eq:E_M-H_E-normalization} \ea
where we identify the model expressions for the electric isovector
$G_E^{(I=1)}(t)$ and magnetic isoscalar $G_M^{(I=0)}(t)$ form factors
\cite{Christov:1995vm,Christov:1995hr}.
    [The factor 3 in the second line of (\ref{Eq:E_M-H_E-normalization})
    appears because for GPDs the notion of (non-) singlet commonly
    refers to quark flavours, e.g.\  $(H^u+H^d)$. In contrast, in
    the case of form factors it refers to proton and neutron,
    e.g. $G^{(I=0)}_M\equiv G^p_M+G^n_M$.]
From (\ref{Eq:form-factor-relations},~\ref{Def:E_M},~\ref{Def:H_E},~\ref{Eq:E_M-H_E-normalization})
and recalling that the leading large-$N_c$ GPDs $(H^u+H^d)(x,\xi,t)$
and $(E^u-E^d)(x,\xi,t)$ are correctly normalized \cite{Petrov:1998kf},
we find for the subleading GPDs in agreement with
Eq.~(\ref{Eq:GPD-form-factor})
\ba
    \int\limits_{-1}^1\di x\,(H^u-H^d)(x,\xi,t)=(F_1^u-F_1^d)(t)\,,\;\;\;
    \int\limits_{-1}^1\di x\,(E^u+E^d)(x,\xi,t)=(F_2^u+F_2^d)(t)\,.
    \label{Eq:check-normalization}
\ea
In App.~\ref{App:polynomiality} it is explicitly demonstrated that the higher
moments in $x$ of $H_E(x,\xi,t)$ and $E_M(x,\xi,t)$ are even polynomials in
$\xi$ according to (\ref{Def:polynom-Hq},~\ref{Def:polynom-Eq}).
In fact, we find that in even moments the coefficients in front of the
highest power in $\xi$ have opposite sign in $(E^u+E^d)(x,\xi,t)$ and
$(H^u+H^d)(x,\xi,t)$ in accordance with the relation (\ref{Eq:relation-h-e}).

\subsection{Spin sum rule}

For the second moment of $E_M(x,\xi,t)$ at $t=0$ we obtain in the $\chi$QSM,
cf.\  App.~\ref{App:proof-of-spin-sum-rule}
\be\label{Eq:spin-sum-rule-ok}
    \int_{-1}^1\di x\;x E_M(x,\xi,0) = 2S^Q + 2L^Q = 2 S^N = 1 \;,\ee
where $S^N=\frac12$ is the total spin of the nucleon, and $S^Q$ and $L^Q$
are the respective contributions of the spin and orbital angular momentum
of quarks and antiquarks to the nucleon spin.
The result in Eq.~(\ref{Eq:spin-sum-rule-ok}) is consistent. In the
effective theory the contribution of quark and antiquark degrees of
freedom, $J^Q=S^Q+L^Q$, must account entirely for the nucleon spin,
since there are no gluons in the model.
Thus, the result in (\ref{Eq:spin-sum-rule-ok}) means that the spin sum rule
(\ref{Eq:spin-sum-rule}) is consistently fulfilled in the model:
\be\label{Ji:XXX2a}
    \int\limits_{-1}^1\!\!\di x\;x(H^u+H^d+E^u+E^d)(x,\xi,0) = 2J^Q = 1
    \;.\ee
We note that in the model $2S^Q=g_A^{(0)}=0.35$ where $g_A^{(0)}$ denotes
the isosinglet axial coupling constant \cite{Goeke:2000wv}.
Thus, in the $\chi$QSM $35\%$ of the nucleon spin are due to the quark spin,
while $65\%$ are due to the orbital angular momentum of quarks and antiquarks.

In a gauge theory it is not possible to separate unambiguously spin and orbital
momentum \cite{Ji:1995cu,Hagler:1998kg}. It also is by no means clear that
the identification of the model expression for quark orbital momentum in
Eq.~(\ref{Eq:spin-sum-rule-ok}) is unambiguous. It is interesting to note
that one arrives at the same result by intuitively identifying
$\hat{L}^i=\epsilon^{ijk}\hat{x}^j\hat{p}^k$
with the orbital angular momentum operator of the effective theory
(\ref{Eq:eff-theory}), cf.\  \cite{Wakamatsu:2000nj}.
However, one has to keep in mind that this issue can rigorously be studied
in the instanton vacuum model from which the $\chi$QSM was derived. Noteworthy
in this context is that $g_A^{(0)}$ in the chiral quark soliton model arises
from the realization of the axial anomaly in the instanton vacuum
\cite{Diakonov:1995qy}.

Since in the chiral quark-soliton model also the total momentum of the nucleon
is carried by quarks and antiquarks only \cite{Diakonov:1996sr}, i.e.\ $M^Q=1$,
we obtain for the second moment of $(E^u+E^d)(x,0,0)$ the result
\be\label{Eq:Eu+Ed-2nd-mom}
    \int\limits_{-1}^1\!\!\di x\;x(E^u+E^d)(x,0,0)=(2J^Q-M^Q)=0\;.
\ee
Interestingly, the result (\ref{Eq:Eu+Ed-2nd-mom}) holds also in QCD in
the asymptotic limit of a large normalization scale $\mu\to\infty$. This
happens because $M^Q$ and $2J^Q$ have the same asymptotics \cite{Ji:1995cu}
\be\label{Eq:MQ-JQ-asymptotic}
    \lim\limits_{\mu\to\infty}M^Q = \lim\limits_{\mu\to\infty}(2J^Q)
    = \frac{3N_F}{16+3N_F}
\ee
where $N_F$ is the number of quark flavours. It should be stressed, however,
that at a low scale $\mu\sim 600\,{\rm MeV}$ in the model $M^Q=2J^Q=1$
is far from the asymptotic values in Eq.~(\ref{Eq:MQ-JQ-asymptotic}).

Eq.~(\ref{Eq:Eu+Ed-2nd-mom}) is the general prediction of any model
lacking explicit gluon degrees of freedom, which is able to consistently
describe the nucleon in terms of quark and antiquark degrees of freedom
-- such as the $\chi$QSM. One can show, using methods of theory of the 
instanton vacuum \cite{Diakonov:1995qy} that the gluon contribution to 
the nucleon momentum and angular momentum is parametrically suppressed by 
the packing fraction of the instantons in the vacuum. Therefore in order 
to obtain non-zero gluon contributions one has to extend $\chi$QSM beyond 
the leading order in the instanton packing fraction.

Noteworthy, it has been argued that momentum and angular momentum should
be equally distributed among quarks and gluons at {\sl any scale}, not
only in the asymptotic limit (\ref{Eq:MQ-JQ-asymptotic}), an observation
which can be reformulated as the absence of an anomalous gravitomagnetic
moment of the nucleon \cite{Teryaev:1999su}.

\subsection{Forward limit}

Finally let us discuss the forward limit. From (\ref{Def:H_E}) we see
that $H_E(x,0,0)=(H^u-H^d)(x,0,0)$. Indeed, taking $\bDelta\to 0$ in
Eq.~(\ref{H_E-model}) and making use of the hedgehog symmetry we recover, in
agreement with Eq.~(\ref{Eq:forward-limit-H}), the model expression for the
flavour non-singlet unpolarized distribution function \cite{Pobylitsa:1998tk}
\be\label{Eq:H_E-forward}
    H_E(x,0,0)
    = \frac{\Mn N_c}{12I}\tripplesum{m,{\rm occ}}{j,{\rm all}}{m\neq j}
    \biggl[\frac{2}{E_j-E_m}-\frac{\partial\;\;}{\partial x\Mn}\biggr]
    \la m|\tau^a|j\ra\,\la j|\,\tau^a(1+\gamma^0\gamma^3)\,
    \delta(x\Mn-E_m-\hat{p}^3)\,|m\ra
    \equiv (f_1^u-f_1^d)(x).\ee

The forward limit of $E_M(x,\xi,t)\to(E^u+E^d)(x,0,0)+(f_1^u+f_1^d)(x)$
contains a contribution which is a priori not known, namely $(E^u+E^d)(x,0,0)$.
Therefore we derive here the model expression for $E_M(x,0,0)$ which we shall
study below in detail. The limit $\xi\to 0$ is regular\footnote{
	Of course, by limiting oneself to $\xi=0$ one drops important physics.
	E.g., the skewedness parameter is related to the correlation length 
	between partons, see \cite{Freund:2002ff} and references therein. 
	The present calculation unfortunately cannot shed any light in this 
	respect.},
however, to complete the forward limit we have to consider with care the 
limit $\bDelta_\perp\to 0$ of the structure
\be\label{forward-lim3}
    \frac{\epsilon^{3jk} \Delta_\perp^j}{\bDelta_\perp^2}
    \exp(i\bDelta_\perp\hat{\bf X}_\perp)
    = \epsilon^{3jk} \;\frac{\Delta_\perp^j}{\bDelta_\perp^2}
    +i\epsilon^{3jk} \hat{X}^m_\perp\;
      \frac{\Delta_\perp^j\Delta_\perp^m}{\bDelta_\perp^2}
    + {\cal O}(\Delta^k) \;. \ee
The first term in the above expansion (for small but non-zero $\bDelta_\perp$)
yields a vanishing result when inserted into Eq.~(\ref{E_M-model}) due to the
hedgehog symmetry. The second term from the expansion in (\ref{forward-lim3})
yields the only contribution which survives the forward limit in
Eq.~(\ref{E_M-model}). Making use of
\be\label{forward-lim5}
    \lim\limits_{\bDelta_\perp\to 0}
    \frac{\Delta_\perp^j\Delta_\perp^m}{\bDelta_\perp^2}
    =\frac{1}{2}\;\delta_\perp^{jm}
    \equiv\frac{1}{2}\;(\delta^{jm}-\delta^{j3}\delta^{m3}) \; \ee
we obtain upon use of hedgehog symmetry
\be\label{forward-lim6}
    E_M(x,0,0) =
        \frac{\Mn^2 N_c}{4I}\;\epsilon^{3jk}
    \tripplesum{m,{\rm occ}}{\;j,{\rm all}}{m\neq j} \biggl[
    \frac{2}{E_j-E_m}-\frac{\partial\;\;}{\partial x\Mn}\biggr]
    \la m|\tau^k|j\ra\,\la j|\,(1+\gamma^0\gamma^3)\,
    \delta(x\Mn-E_m-\hat{p}^3)\,\hat{X}_\perp^j\,|m\ra\;.\ee
As a final check we integrate $E_M(x,0,0)$ in Eq.~(\ref{forward-lim6}) over
$x$ and recover the model expression for the isoscalar magnetic moment
\cite{Wakamatsu:1990ud}
\be\label{chek-magn-mom-1}
        \int\limits_{-1}^1\!\!\di x\,E_M(x,0,0)
    = \frac{\Mn N_c}{2I}\;\doublesum{m,{\rm occ}}{\;j,{\rm non}}
        \frac{1}{E_j-E_m}\, \la m|\tau^3|j\ra\, \la j|\,\gamma^0
    (\bgam \times \hat{\bf X})^3\,|m\ra = 3\mu^{(T=0)} \;.\ee

  \section{\boldmath The forward limit of $(E^u+E^d)(x,\xi,t)$}
  \label{Sec:6-Eu+Ed-forward-lim}

In order to compute the forward limit
$(E^u+E^d)(x,0,0)\equiv E_M(x,0,0)-(f_1^u+f_1^d)(x)$ we have to
numerically evaluate the model expression (\ref{forward-lim6}) for
$E_M(x,0,0)$. The distribution function $(f_1^u+f_1^d)(x)$ was already
studied in \cite{Diakonov:1996sr,Diakonov:1997vc,Weiss:1997rt}.
The numerical methods needed for that were developed in
Refs.~\cite{Diakonov:1997vc,Pobylitsa:1998tk,Weiss:1997rt} in the context of
usual parton distributions functions in leading and subleading order of the
large-$N_c$ expansion. We restrict ourselves to the forward limit since this
numerical technic does not allow a full calculation of GPDs for all values
of $\xi$ and $t$.

  \subsection{Gradient expansion and chiral enhancement of GPDs }
  \label{App:gradient-expansion}

Before studying the model expression (\ref{E_M-model}) numerically in the
model we shall consider the gradient expansion of $E_M(x,\xi,t)$.
This expansion consists in expanding model expressions in powers
of the gradients of the (static) chiral field $\nabla U$.
Such an expansion would quickly converge if the soliton field were slowly
varying, i.e.\ if the soliton were large $\nabla U\sim 1/R_{\rm sol}\ll M$
where $R_{\rm sol}$ is the scale characterizing the soliton size. However,
for the physical soliton solution $R_{\rm sol}\sim 1/M$.
Nevertheless the gradient expansion is instructive and allows --
among others -- to study the UV-behaviour of the model expressions.
Also as we shall see below the gradient expansion allows to obtain
strong enhancement of GPD $E_M(x,\xi,t)$ in the region of small $x$. 
This enhancement is related to the effect of the pion cloud.

\begin{figure}[t!]
        \includegraphics[width=5.4cm,height=5.2cm]{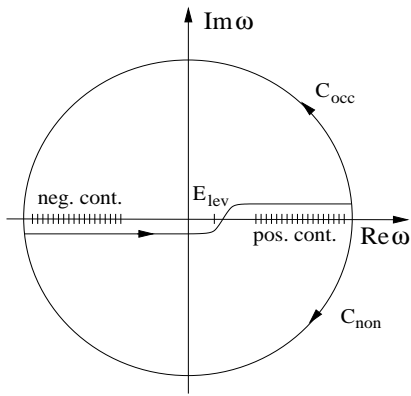}
    \caption{\label{Fig:1-contour}\footnotesize
    The contour $C$ of the $\omega$-integral in Eq.~(\ref{Eq:grad-00}).}
\end{figure}

The model expression for $E_M(x,\xi,t)$ can be rewritten equivalently
in the following way
\ba\label{Eq:grad-00}
    E_M(x,\xi,t)
&=&
    \frac{N_c\Mn^2\epsilon^{3bc}\Delta^b}{2I\,\bDelta^2_\perp}\;
        {\rm tr}_{\gamma,F}\int_C\frac{\di\omega\di^3{\bf p}}{(2\pi)^4}
    \la{\bf p}-\fracS{\bDelta}{2}|
    \nonumber\\
&&  \times\Biggl[
        \delta(\omega+p^3-x\Mn)(1+\gamma^0\gamma^3)\,
    \frac{1}{\omega-H}\,\tau^c\,\frac{1}{\omega-H}
    \nonumber\\
&&  \;\,+\;
    \frac{\partial}{\partial x\Mn}\;\delta(\omega+p^3-x\Mn)
    (1+\gamma^0\gamma^3)\tau^c\frac{1}{\omega-H}
    \Biggr]|{\bf p}+\fracS{\bDelta}{2}\ra
\ea
where ``${\rm tr}_{\gamma,F}$'' denotes the trace over Dirac- and flavour
indices. The contour $C$ is defined in Fig.~\ref{Fig:1-contour}. Closing 
the contour $C$ in the upper half of the complex $\omega$-plane yields the 
expression in Eq.~(\ref{E_M-model}), closing it in the lower half plane one 
obtains the equivalent expression where the summation goes over non-occupied 
states $m$, cf.\  the sequence of Eq.~(\ref{E_M-model}).

Expanding the expression in Eq.~(\ref{Eq:grad-00}) in powers of $\nabla U$
one obtains\footnote{
    For that we rewrite the operator
    $1/(\omega-H)=\frac12(\omega+H)/(\omega^2-H^2)+$
    $\frac12(\omega^2-H^2)^{-1}(\omega+H)$ where
    $H^2=\hat{\bf p}^2+M^2+iM\bgam\bnabla U^{\gamma_5}$,
    and expand it in powers of $iM\bgam\bnabla U^{\gamma_5}$.
    Such a ``symmetric expansion'' of $1/(\omega-H)$ ensures the
    hermiticity of the operator also in the case when the series in
    $\nabla U$ is truncated.}
already from the zeroth order in $(\nabla U)$ the following result for the
continuum contribution
\ba\label{Eq:grad-01}
    E_M^{\rm cont}(x,\xi,t)
&=& \frac{\Mn^2 f_\pi^2}{4I\,\bDelta_\perp^2}\int\frac{\di^3{\bf k}}{(2\pi)^3}
    f({\bf k}+\fracS{\bDelta}{2})f({\bf k}-\fracS{\bDelta}{2})
    \biggl[-\bDelta^2\,k^3+\Delta^3(\bDelta{\bf k})\biggr]\nonumber\\
&&  \times\Biggl\{
    \frac{\Theta(x\Mn+k^3)\Theta(\xi-x)-\Theta(-x\Mn-k^3)\Theta(x-\xi)}
         {k^3-\xi\Mn}  + (\xi\leftrightarrow -\xi) \nonumber\\
&&  \;\;+
    \frac{\Theta(x-\xi)\Theta(-x-\xi)-\Theta(-x+\xi)\Theta(x+\xi)}{2\xi\Mn}
    \Biggr\}+{\cal O}(\nabla U)
\ea
where $f({\bf q})=f(|{\bf q}|)$ is one of the functions describing the
Fourier transform of the $U({\bf x})$-field
\be\label{Eq:grad-02}
    \int\di^3{\bf x}\;(U({\bf x})-1)\,e^{-i\bf qx}
    =g({\bf q})+\frac{i\btau\bf q}{|{\bf q}|}f({\bf q})
    \;;\;\;\;
    f({\bf q})\equiv 4\pi\int_0^\infty\di r\;r^2j_1(r|{\bf q}|)\;\sin P(r)
    \;.\ee
Note that the property $E_M(x,\xi,t)=E_M(x,-\xi,t)$ holds manifestly for the
result in Eq.~(\ref{Eq:grad-01}).

In the derivation of Eq.~(\ref{Eq:grad-01}) we encountered a logarithmic
UV-divergence proportional to $M^2$. This UV-divergence can be regularized,
e.g., by means of a Pauli-Villars subtraction as
\be\label{Eq:grad-03}
    E_M^{\rm cont}(x,\xi,t)_{\rm reg}=
    E_M^{\rm cont}(x,\xi,t;M)-\frac{M^2}{M_{\rm PV}^2}
    E_M^{\rm cont}(x,\xi,t;M_{\rm PV}).
\ee
We removed the dependence of the result on cutoff and regularization 
scheme in favour of a physical parameter, namely the pion decay constant 
$f_\pi=93\,{\rm MeV}$ given in the effective theory (\ref{Eq:eff-theory}) 
by the (Euclidean loop) integral
\be\label{Eq:grad-04}
    f_\pi^2= \int\frac{\di^4 p_E}{(2\pi)^4}\;
    \frac{4N_cM^2}{(p_E^2+M^2)^2}\Biggl|_{\rm reg}
\ee
which exhibits a similar UV-behaviour and can be regularized analogously.
In the first order of the gradient expansion, indicated only symbolically
as ${\cal O}(\nabla U)$ in Eq.~(\ref{Eq:grad-01}), we also encounter a
logarithmic divergence which can be removed by means of (\ref{Eq:grad-03}).
Still higher orders in the gradient expansion yield finite results --
as one can conclude from dimensional counting. (Or from the fact that
the isoscalar magnetic moment appears at the order $(\nabla U)^2$ and is
UV-finite, cf. Ref.~\cite{Diakonov:1987ty}.)

Eq.~(\ref{Eq:grad-01}) may give at best only a rough estimate for
$E_M(x,\xi,t)$. Nevertheless it contains already main features of the
total result. Apart from the UV-properties, which we discussed above,
the small $x$-behaviour in the chiral limit is of interest since the
exact numerical calculation is performed under such conditions.
The soliton profile behaves at large distances as
\be\label{Eq:grad-05}
    P(r)= -\frac{B}{r^2}\,(1+m_\pi r)\,\exp(-m_\pi r)\;,\;\;\;
    B =\frac{3\,g_A^{(3)}}{8\pi \,f_\pi^2}\;\;\;\mbox{for}\;\;
    r\gg \sqrt{B} \ee
with $B$ related to axial isovector coupling constant $g_A^{(3)}=1.26$ by
means of equations of motion. The large distance behaviour (\ref{Eq:grad-05})
of the soliton profile translates into the infrared-behaviour of
the function $f({\bf q})$ as
\be\label{Eq:grad-05a}
    f({\bf q})=
    -\,\frac{3\,g_A^{(3)}}{2f_\pi^2}\;\frac{1}{{\bf q}^2+m_\pi^2}
    \;\;\;\mbox{for small ${\bf q}^2,m_\pi^2\ll B^{-1}$.}
\ee
Note that $B^{-1} < (4\pi\,f_\pi)^2\sim 1\,{\rm GeV}^2$ which is a typical
scale for chiral symmetry breaking effects.
Apparently the limits $m_\pi\to 0$ and ${\bf q}\to 0$ do not commute.\footnote{
	The non-commutativity of such limits is known, e.g., from studies
	of the polarized photon structure function, where the virtuality 
	of the photon plays the role of the momentum ${\bf q}$. In that
	case it is known that the correct order of limits is to take
	first the photon virtuality to zero, and only then to consider
	$m_\pi\to 0$ \cite{Freund:1994ti}. However, in our context it 
	is not clear in which order the limits should be taken, though
	phenomenology may suggest first to keep $m_\pi$ finite.}
In the forward limit in Eq.~(\ref{Eq:grad-01}) this means
\be\label{Eq:grad-6}
    E_M(x,0,0)=\frac{\Mn^2f_\pi^2}{6I(2\pi)^2}\int\limits_{|x\Mn|}^\infty
    \!\!\!\!\!\di q\; q^2 f(q)^2\biggl(1-\frac{|x\Mn|}{q}\biggr)
    \stackrel{x\to 0}{=}
    \biggl[\frac{3 g_A^{(3)}}{8\pi f_\pi}\biggr]^2\;\frac{\Mn}{I} \cdot
    \cases{
    {\displaystyle{\frac{1}{x}}}   & for $m_\pi = 0$, \cr
    \phantom{X} & $\phantom{X}$ \cr
    {\displaystyle{\frac{\pi}{2}\,\frac{M_N}{m_\pi}}} & for $m_\pi\neq 0$.}
\ee
Note that the soliton moment of inertia could be eliminated in favour
of the $\Delta$-nucleon mass-splitting as $M_\Delta-\Mn = 3/(2I)$.

 Although
derived in the leading order of the gradient expansion of model expressions Eq.~(\ref{Eq:grad-6})
is of more general nature as it is related to the effect of pion cloud. Therefore it is should be possible
to derive the above expression by means of chiral perturbation theory. Another interesting question is to find
phenomenological manifestations of the chiral enhancement. This would open new exciting possibilities to
investigate chiral symmetry breaking in the hard exclusive processes.

\subsection{Numerical calculation}

As any quantity in the model, $E_M(x,0,0)$ is composed of a contribution
of the discrete level and of the continuum contribution defined,
respectively, as
\ba
    E_M^{\rm lev}(x,0,0)
    =
    \frac{\Mn^2 N_c}{4I}\;\epsilon^{3jk}
    \doublesum{E_j,{\rm all}}{j\neq {\rm lev}} \biggl[
    \frac{2}{E_j-E_{\rm lev}}-\frac{\partial\;\;}{\partial x\Mn}\biggr]
    \la{\rm lev}|\tau^k|j\ra\,\la j|\,(1+\gamma^0\gamma^3)\,
    \delta(x\Mn-E_m-\hat{p}^3)\,\hat{X}_\perp^j\,|{\rm lev}\ra
    \!\!\!\!\!  &&  \!\!\!\!\! \nonumber\\
    E_M^{\rm cont}(x,0,0)
    =
    \frac{\Mn^2 N_c}{4I}\;\epsilon^{3jk}
    \tripplesum{E_m<0}{E_j,{\rm all}}{m\neq j} \biggl[
    \frac{2}{E_j-E_m}-\frac{\partial\;\;}{\partial x\Mn}\biggr]
    \la m|\tau^k|j\ra\,\la j|\,(1+\gamma^0\gamma^3)\,
    \delta(x\Mn-E_m-\hat{p}^3)\,\hat{X}_\perp^j\,|m\ra
     \!\!\!\!\! && \!\!\!\!\!  \nonumber\\
    = -
    \frac{\Mn^2 N_c}{4I}\;\epsilon^{3jk}
    \tripplesum{E_m>0}{E_j,{\rm all}}{m\neq j} \biggl[
    \frac{2}{E_j-E_m}-\frac{\partial\;\;}{\partial x\Mn}\biggr]
    \la m|\tau^k|j\ra\,\la j|\,(1+\gamma^0\gamma^3)\,
    \delta(x\Mn-E_m-\hat{p}^3)\,\hat{X}_\perp^j\,|m\ra
    \!\!\!\!\!  && \!\!\!\!\! \nonumber\\
    \label{Eq:EM-lev-cont}\ea
For a constant (momentum-independent) $M$ the continuum contribution to
$E_M(x,\xi,t)$ is UV-divergent. It can be regularized by means of a
single Pauli-Villars according to Eq.~(\ref{Eq:grad-03}).
In the context of parton distribution functions the Pauli-Villars method
is the preferred regularization scheme because it preserves fundamental
properties of parton distributions such as sum rules, positivity,
etc.\  \cite{Diakonov:1996sr}. It should be noted that the contribution of
the discrete level, which is always finite, must not be regularized
\cite{Weiss:1997rt}.

The numerical method to evaluate $E_M(x,0,0)$ consists in placing the
soliton in a large but finite spherical box, introducing free basis states
(eigenstates of the free Hamiltonian, see Eq.~(\ref{Eq:Hamiltonian}) and
below), and discretizing the basis by imposing appropriate boundary conditions
following Kahana and Ripka \cite{Kahana:1984be}.
The full Hamiltonian (\ref{Eq:Hamiltonian}) can be diagonalized numerically in
this basis \cite{Wakamatsu:1990ud}. With the eigenfunctions and eigenstates,
$\Phi_n({\bf x})$ and $E_n$, one is then in a position to evaluate
Eq.~(\ref{Eq:EM-lev-cont}). This is most conveniently done by converting
the model expression (\ref{Eq:EM-lev-cont}) into a spherically symmetric
form \cite{Diakonov:1997vc}.
For that we introduce in (\ref{Eq:EM-lev-cont}) a unit vector ${\bf a}$
such that the original expression is recovered for ${\bf a}={\bf e}^{(3)}$,
\be\label{Eq:average-1}
    \epsilon^{3jk}(1+\gamma^0\gamma^3)\,\delta(x\Mn-E_m-\hat{p}^3)\,
    \hat{X}_\perp^j\, \to
    (1+\gamma^0 {\bf a}\cdot\bgam)\,
    \delta(x\Mn-E_m-{\bf a}\cdot\hat{\bf p})\,
    ({\bf a}\times\hat{\bf X})^k\;.
\ee
Then we average over all possible orientations of the vector ${\bf a}$.
This yields
\ba\label{Eq:average-2}
        E_M(x,0,0)
    &=&
        -\frac{\Mn^2 N_c}{4I}\tripplesum{m,{\rm occ}}{\;j,{\rm all}}{m\neq j}
        \biggl[\;\frac{2}{E_m-E_j}+\frac{\partial\;\;}{\partial x\Mn}\;\biggr]
        \la m|\tau^k|j\ra\,\la j|\Biggl[
    (\hat{\bf p}\times\hat{\bf X})^k\;
        \frac{I_1(|\hat{\bf p}|,E_m,x)}{|\hat{\bf p}|^2}\nonumber\\
    &&
    + (\gamma^0\bgam\times\hat{\bf X})^k\;\frac{(I_0-I_2)
      (|\hat{\bf p}|,E_m,x)}{2|\hat{\bf p}|}
    + (\hat{\bf p}\times\hat{\bf X})^k\,
      (\gamma^0\bgam\cdot\hat{\bf p})\,\frac{(3I_2-I_0)
      (|\hat{\bf p}|,E_m,x)}{2|\hat{\bf p}|^3}\Biggr]\,|m\ra\;,\nonumber\\
  &&    \mbox{where} \;\;\;
    I_l(|\hat{\bf p}|)=\frac{(x\Mn-E_m)^l\!\!}{2\,|\hat{\bf p}|^l}\,
    \Theta(|\hat{\bf p}| -|x\Mn-E_m|)\;.
\ea
Although non-commuting operators appear in (\ref{Eq:average-2}) nevertheless
the final result is a hermitian operator. The averaging procedure must, of
course, preserve hermiticity. It is this step which prevents us from tackling
also the problem to compute numerically $E_M(x,\xi,t)$ for non-zero $\xi$ and
$t$. For finite $\bDelta$ the appearance of an additional direction, namely
along $\bDelta_\perp$ perpendicular to the 3-direction, makes the resulting
expressions unsuitable for a numerical evaluation.

Since we work with a discrete basis, the $\Theta$-functions in
(\ref{Eq:average-2}) would lead to discontinuous functions. Therefore
we convolute the true (but due to finite size effects discontinuous)
function with a narrow Gaussian
\be\label{Eq:smear}
    E_M^{sm}(x,0,0) = \frac{1}{\sqrt{\pi}\gamma}
    \int\limits_{-\infty}^\infty\di x^\prime
    \;e^{-(x-x^\prime)^2/\gamma^2}\;E_M(x^\prime,0,0)\;.
\ee
The width $\gamma$ has to be chosen adequately according to the spacing of
the discrete basis states \cite{Diakonov:1997vc}. The effect of smearing can
be removed by a deconvolution at the end of the calculation.

\begin{figure}[t!]
\begin{tabular}{cc}
    \includegraphics[width=5.4cm,height=5.4cm]{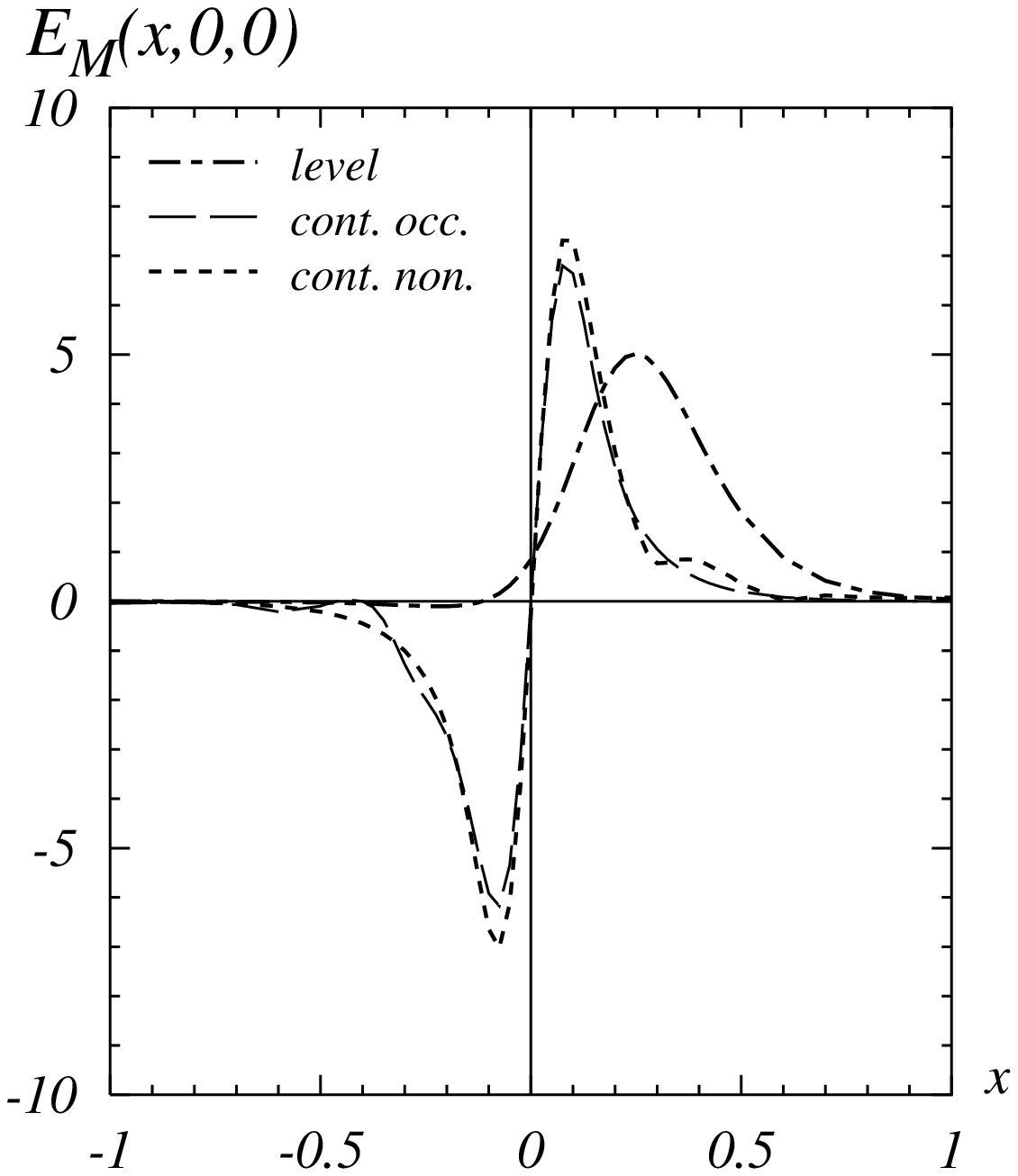}&
    \includegraphics[width=5.4cm,height=5.4cm]{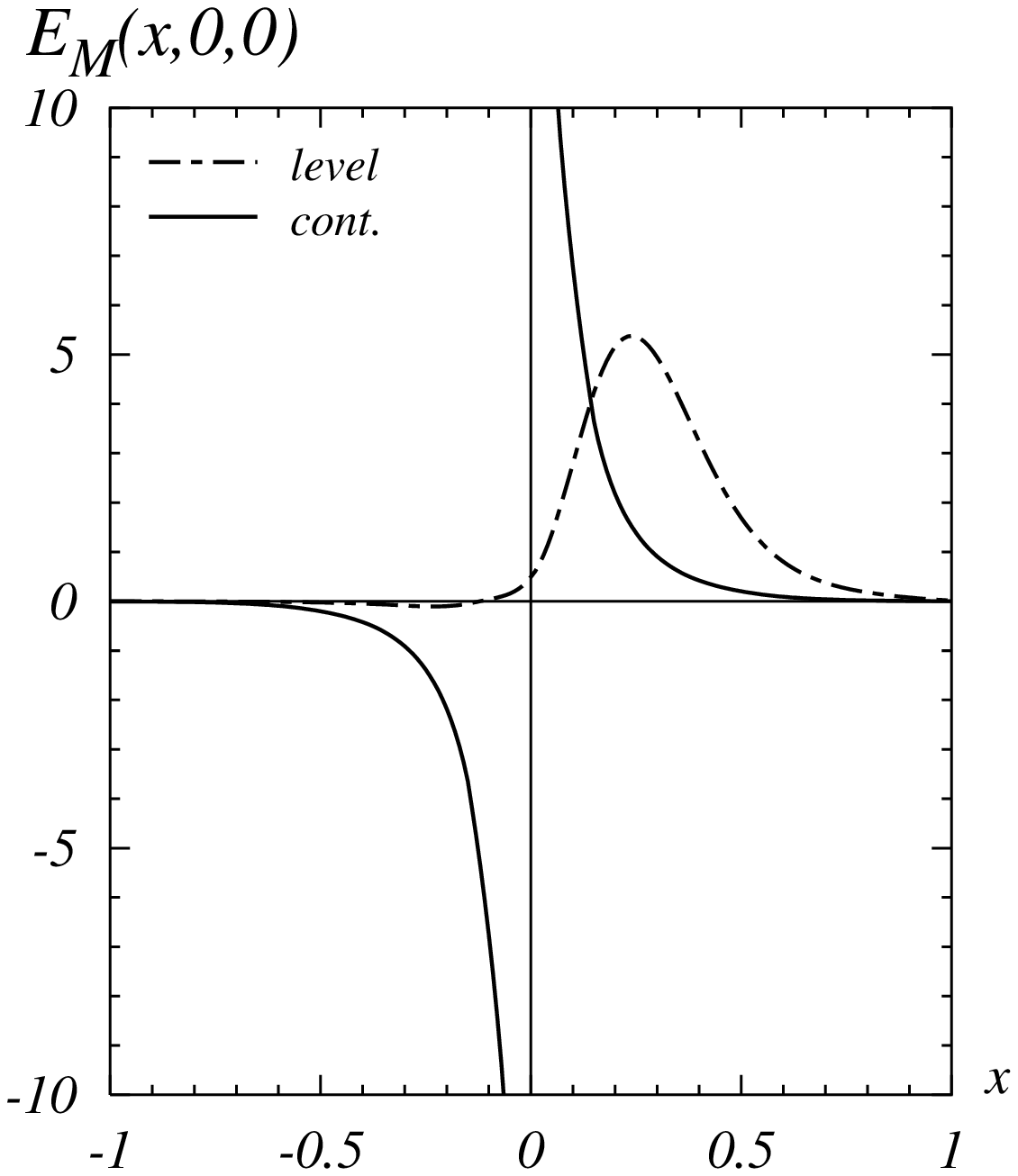}\cr
    \hspace{0.8cm} {\bf a} &
    \hspace{0.8cm} {\bf b}
\end{tabular}
    \caption{\footnotesize\label{Fig:2-EM}
    $E_M(x,0,0)=(E^u+E^d+H^u+H^d)(x,0,0)$ as function of $x$ from the
    $\chi$QSM at a low scale of about $\mu\approx 600\,{\rm MeV}$.
    At negative $x$ the curves show minus the antiquark distributions.
    Fig.~\ref{Fig:2-EM}a shows the (smeared) discrete level and continuum 
    contribution. The latter is evaluated in the two equivalent ways in
    Eq.~(\ref{Eq:EM-lev-cont}), namely by summing respectively over
    occupied and non-occupied states.
    In Fig.~\ref{Fig:2-EM}b the effect of the Gaussian smearing is removed.
    These are the final results for respectively the discrete level
    and the continuum contribution to $E_M(x,0,0)$.}
\end{figure}

Fig.~\ref{Fig:2-EM}a shows the level and continuum contribution to 
$E_M(x,0,0)$ computed with $M=350\,{\rm MeV}$ in a box of the size 
$D_{\rm box}=18\,{\rm fm}$ with $\gamma=0.08$. 
The continuum contribution is evaluated in the two equivalent ways, 
Eq.~(\ref{Eq:EM-lev-cont}). This explicit check of the equivalence of
summations over respectively occupied and non-occupied states is an important
check of the numerical method, and provides a measure for the numerical
accuracy. As can be seen from Fig.~\ref{Fig:2-EM}a the two calculations 
yield the same result within few percent.

We observe that the equivalence is spoiled by unphysical boundary effects
in the chiral limit, when the soliton profile $P(r)\propto 1/r^2$ at large
distances. In order to restore the equivalence it is necessary that the
profile decays faster at large $r$, e.g., as
$P(r)\propto\exp(-m_\pi^\prime r)(1+m_\pi^\prime r)/r^2$ 
(the numerical parameter $m_\pi^\prime$ is not to be confused with the
physical pion mass discussed above).
Since the issue of computing self-consistent profiles with finite pion masses
in the Pauli-Villars regularization is not yet solved, cf.\
\cite{Kubota:1999hx} for a discussion, we use the chiral self-consistent
profile computed in \cite{Weiss:1997rt} up to some $r_A$ and continue it
for $r>r_A$ with an artificial Yukawa-tail-like suppression of the above kind.
We find the results practically independent of $r_A$ and $m_\pi^\prime$ in the range
$r_A=(4-8)\,{\rm fm}$ and $m_\pi^\prime=(100-200)\,{\rm MeV}$. This proves that what
matters in this context is only a sufficiently small value of the profile at
the boundary $r=D_{\rm box}$, and makes the the extrapolations $r_A\to\infty$
and/or $m_\pi^\prime\to 0$ superfluous, which in principle would be necessary to remove
any dependence of these numerical parameters.
The equivalence demonstrated in Fig.~\ref{Fig:2-EM}a is achieved in this way.
This procedure was applied successfully to cure an analogue problem in
the calculation of the transversity distribution \cite{Schweitzer:2001sr}.

Let us comment on the effect of smearing, which is negligible whenever one
deals with a continuous function such as the contribution of the discrete 
level, cf.\  Fig.~\ref{Fig:2-EM}a and \ref{Fig:2-EM}b.
However, in the case of the continuum contribution the smearing ``hides''
an $1/x$ singularity which becomes apparent only in the final result
after the smearing is removed. This is done by Fourier transforming
$E_M^{sm}(x,0,0)$ of Eq.~(\ref{Eq:smear}), dividing out the smearing Gaussian,
and re-Fourier transforming -- which yields the final results shown in 
Fig.~\ref{Fig:2-EM}b.

We remark that in the $\chi$QSM the parton distribution functions (and
forward limits of GPDs) do not vanish for $|x|\ge 1$. Instead they decay
as $\exp(-{\rm const}\,N_c x)$ at $|x|\ge 1$ \cite{Diakonov:1996sr} which is,
in fact, numerically very small even for $N_c=3$.

\subsection{Discussion of the results}

We observe in $E_M(x,0,0)$ that the discrete level contributes predominantly
to the distribution of quarks rather than antiquarks. The continuum, however,
contributes nearly equal portions to quark and antiquark distributions. As a
result the isoscalar magnetic moment receives a negligible contribution from
the continuum. This strong dominance of the level contribution in the
isoscalar magnetic moment was observed in earlier calculations
\cite{Christov:1995hr}. The result (in units of the nuclear magneton)
\be\label{Eq:discuss-01}
    \mu^{(T=0)}\equiv (\mu^p+\mu^n) = \frac13
    \int\limits_{-1}^1\!\!\di x\;E_M(x,0,0) = 0.65
    \;\;\;{\rm vs.}\;\;\; 0.88_{\rm exp}\;,
\ee
agrees with the experimental value to within $25\%$, i.e.\  to within
an accuracy typical for $\chi$QSM results \cite{Christov:1995vm}.

\begin{figure}[t!]
\begin{tabular}{ccc}
    \includegraphics[width=5.4cm,height=5.4cm]{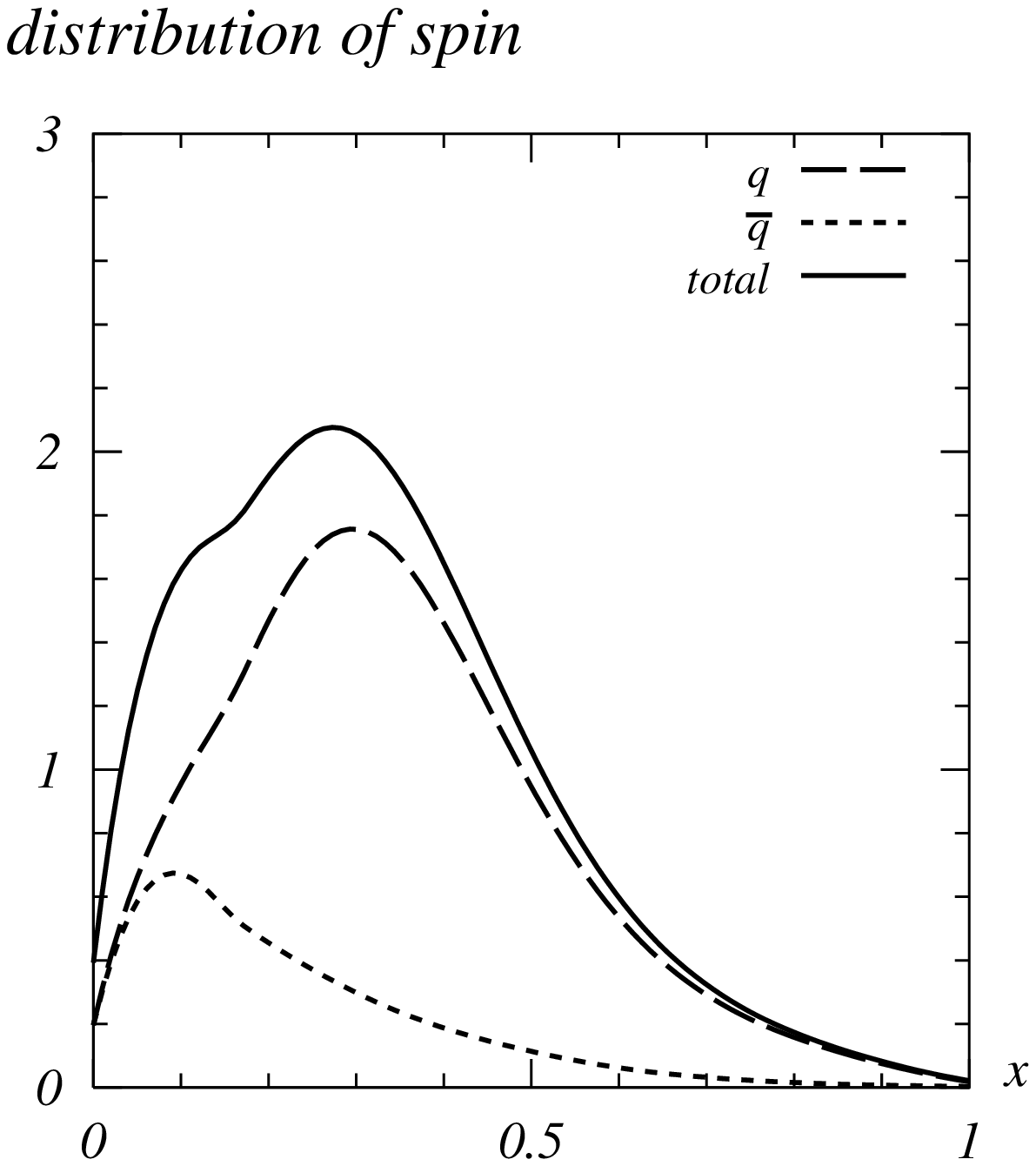}&
    \includegraphics[width=5.4cm,height=5.4cm]{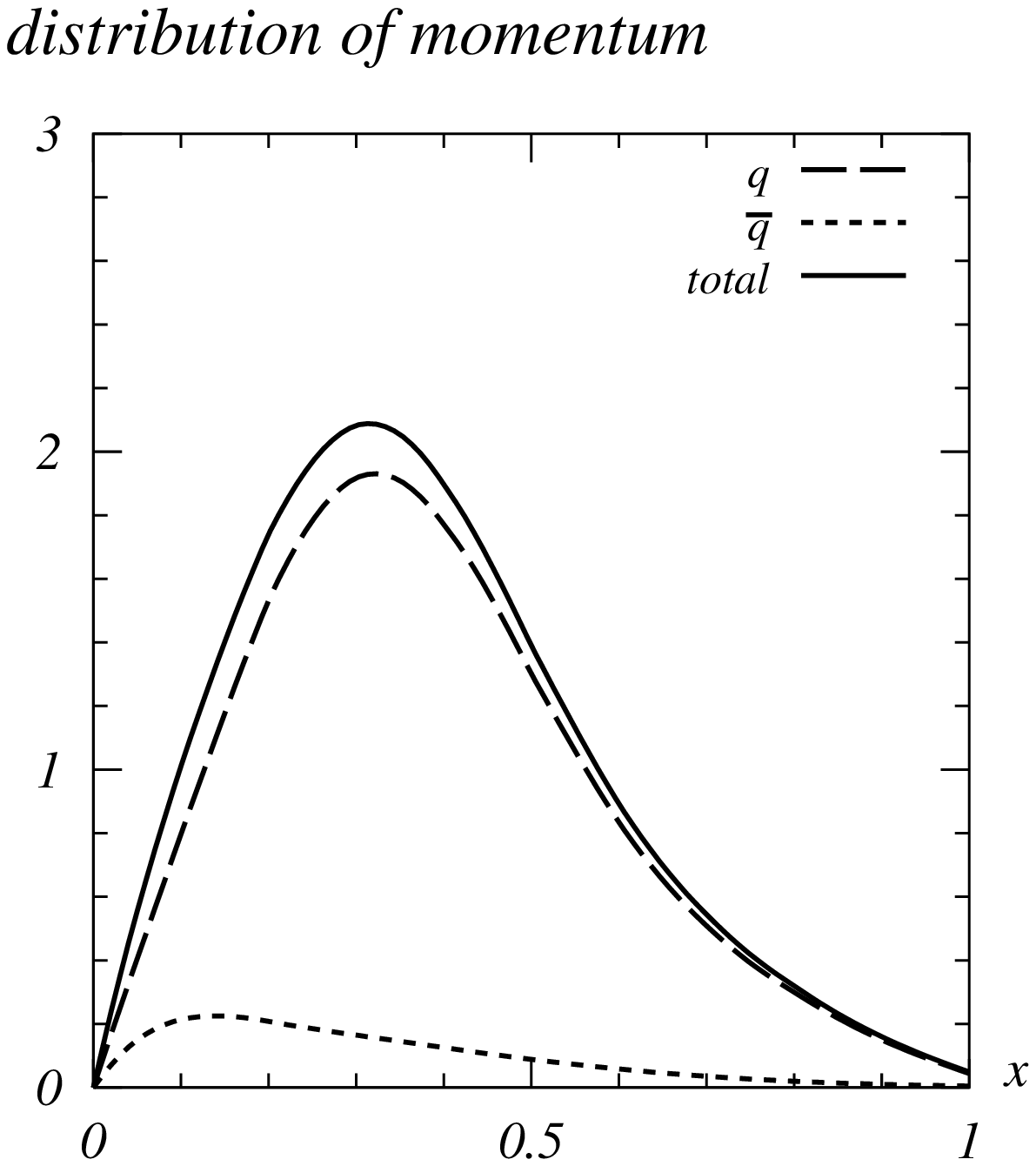}&
    \includegraphics[width=5.4cm,height=5.4cm]{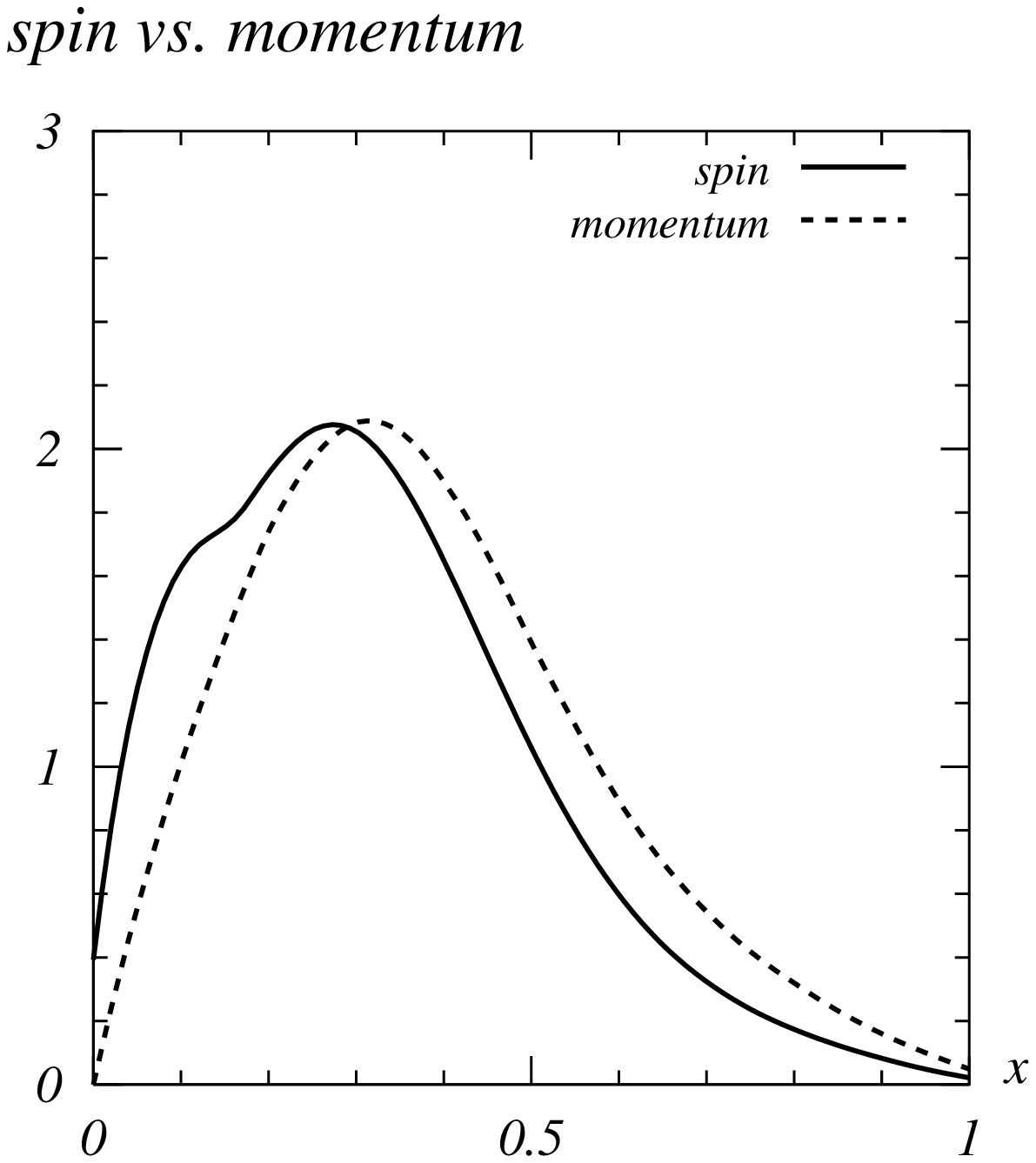}\cr
    \hspace{0.8cm} {\bf a} &
    \hspace{0.8cm} {\bf b} &
    \hspace{0.8cm} {\bf c}
\end{tabular}
    \caption{\footnotesize\label{Fig:3-spin-mom}
    Comparison of the ``spin'' and ``momentum distribution'' in the
    nucleon in the $\chi$QSM at a low scale of about $\mu=600\,{\rm MeV}$.
    Fig.~\ref{Fig:3-spin-mom}a shows the respective contributions of quarks and antiquarks
    to the nucleon spin as distributions in $x$, defined as
    $xE_M(x,0,0)=x(E^u+E^d+H^u+H^d)(x,0,0)$ (and analog for antiquarks).
    The sum of the integrals over $x$ of the quark and antiquark
    contributions yields $2J^Q=1$.
    Fig.~\ref{Fig:3-spin-mom}b shows the distribution of the momentum in the nucleon,
    $x(f_1^u+f_1^d)(x)$, for quarks (and analog for antiquarks).
    The sum of the integrals over $x$ of the quark and antiquark
    contributions yields $M^Q=1$.
    In Fig.2c the total spin and momentum distributions are compared.}
\end{figure}

We can interpret $xE_M(x,0,0)\equiv x(H^u+H^d+E^u+E^d)(x,0,0)$ (and analog
for antiquarks) as the distribution of spin in the nucleon in the following
sense. This quantity tells us how much of the nucleon spin is due to quarks
(and antiquarks) from which region in $x$. In Fig.~\ref{Fig:3-spin-mom}a 
we see that the contribution due to quarks is concentrated around
$x\approx 0.3$, while the contribution of antiquarks has its maximum in the
region $x\approx 0.1$. It is worthwhile stressing that a sizeable fraction of
the nucleon spin is due to antiquarks already a the low scale of the model.
We find the nucleon spin distributed among antiquarks and quarks as
$\mbox{antiquarks}:\mbox{quarks}=1:4$.  Integrating $xE_M(x,0,0)$ over $x$
and summing up the contributions of quarks and antiquarks we obtain unity
(within $2\%$ numerical accuracy) -- in agreement with the spin sum rule
in the model, Eq.~(\ref{Ji:XXX2a}).

It is instructive to compare -- within the model -- the spin distribution
to the momentum distribution of quarks and antiquarks in the nucleon.
First we note that antiquarks are sizeable also in the momentum distribution,
however, less pronounced than in the case of the spin distribution,
see Fig.~\ref{Fig:3-spin-mom}b. The nucleon momentum is distributed as
$\mbox{antiquarks}:\mbox{quarks}\approx 1:10$. Parameterizations
performed at comparably low scales \cite{Gluck:1998xa} indicate that about
$30\%$ of the nucleon momentum is carried by gluons already at low scales.
Thus, the distribution of momentum in the model seems to overestimate the
contribution of the quarks and antiquarks by about $30\%$, which is within the
expected accuracy of the model \cite{Diakonov:1998ze}.
It will be interesting to see whether the situation is similar in the case
of the spin distribution. 

Finally, comparing the momentum and spin distributions, 
Fig.~\ref{Fig:3-spin-mom}c, we observe that the spin distribution is 
somehow softer, i.e.\ shifted towards smaller $x$.

\begin{figure}[t!]
\begin{tabular}{ccc}
    \includegraphics[width=5.4cm,height=5.4cm]{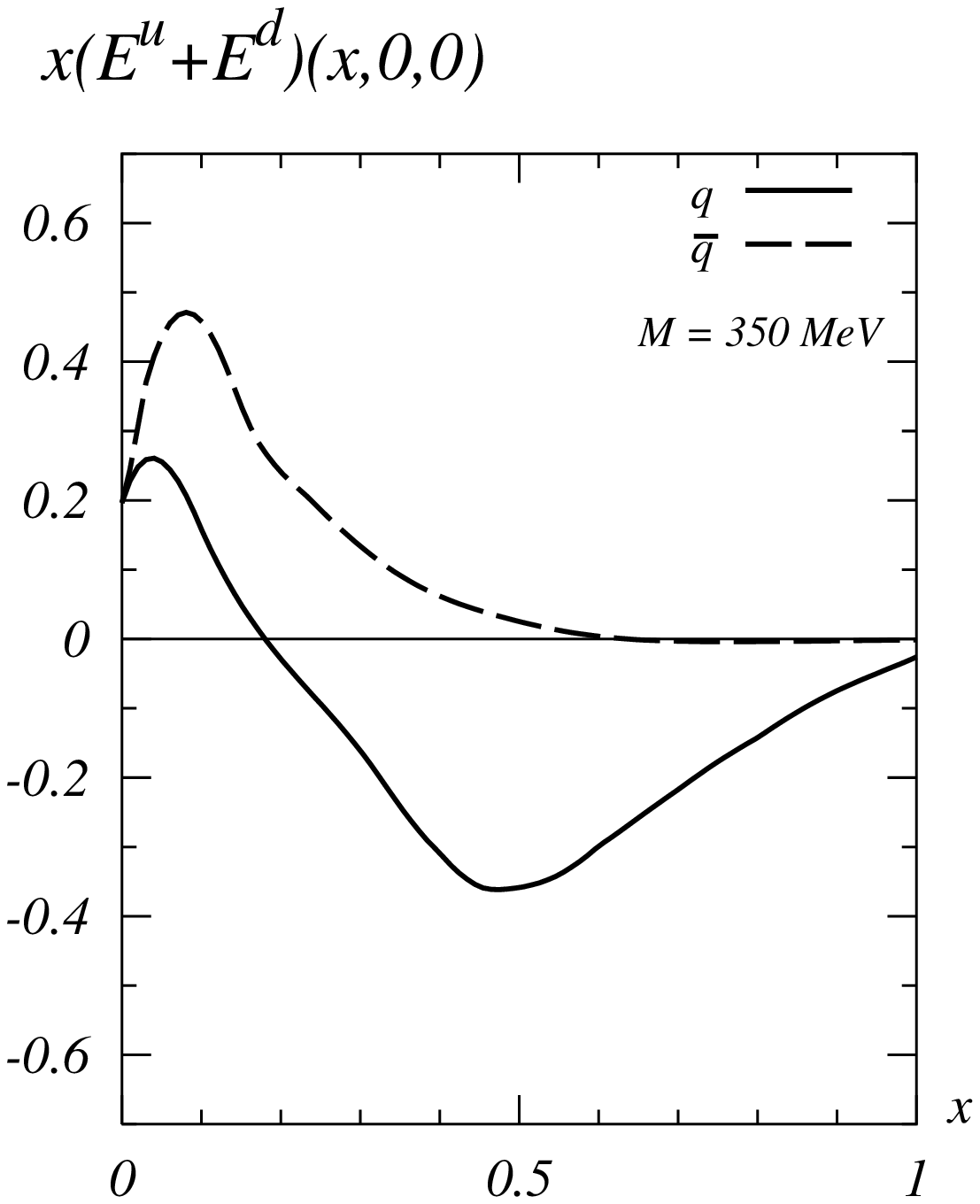}&
    \includegraphics[width=5.4cm,height=5.4cm]{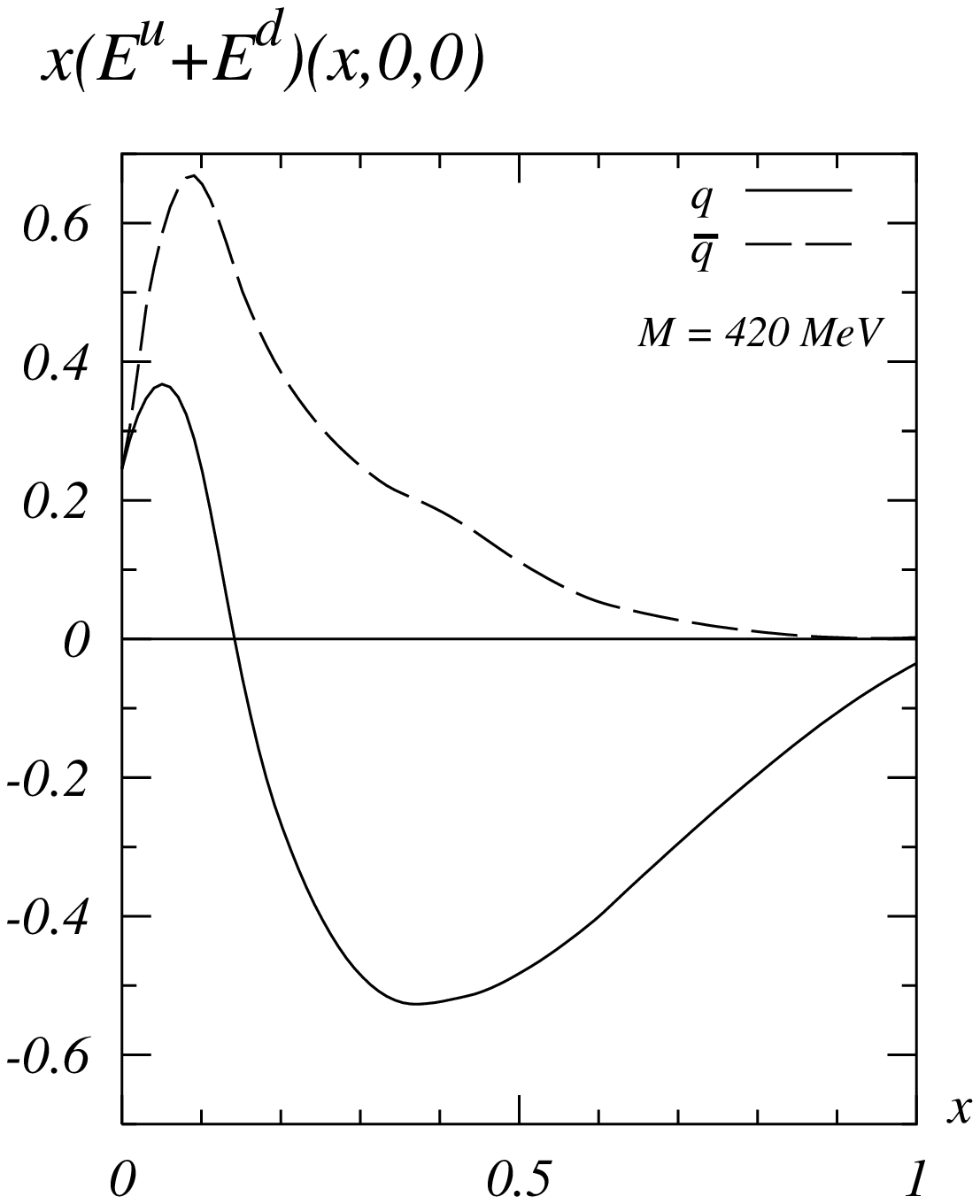}&
    \includegraphics[width=5.4cm,height=5.4cm]{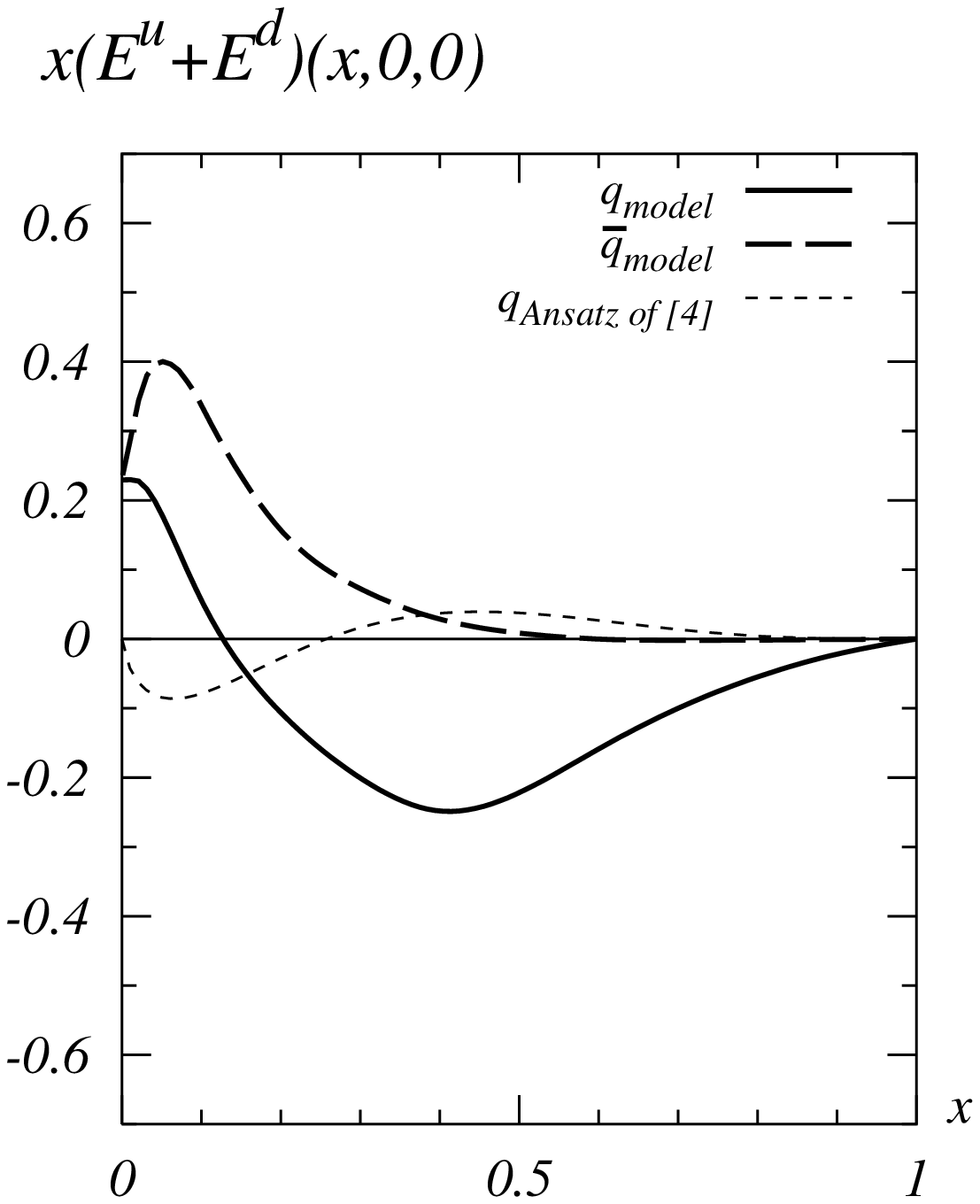}
    \cr
        \hspace{0.8cm} {\bf a} &
    \hspace{0.8cm} {\bf b} &
    \hspace{0.8cm} {\bf c}
\end{tabular}
    \caption{\footnotesize\label{Fig:4-Eu+Ed-final}
    Fig.~\ref{Fig:4-Eu+Ed-final}a shows the final result for $x(E^u+E^d)(x,0,0)$ and
    $x(E^{\bar u}+E^{\bar d})(x,0,0)$. For sake of comparison,
    see text, Fig.~\ref{Fig:4-Eu+Ed-final}b shows the result one obtains with a
    constituent mass $M=420\,{\rm MeV}$.
    Fig.~\ref{Fig:4-Eu+Ed-final}c shows the final model result (Fig.~\ref{Fig:4-Eu+Ed-final}a) LO-evolved from the
    initial scale taken as $\mu^2=0.36\,{\rm GeV}^2$ to the scale
    $Q^2=5\,{\rm GeV}^2$ typical in experiments. For comparison the
    Ansatz of Ref.~\cite{Goeke:2001tz} is shown for quarks (dotted line).
    The corresponding antiquark distribution was assumed to be zero in
    Ref.~\cite{Goeke:2001tz}. Note that the effect of sea quarks was
    simulated in \cite{Goeke:2001tz} by a $\delta$-function not visible
    in the plot.}
\end{figure}

In Fig.~\ref{Fig:4-Eu+Ed-final}a, we present the final results for $x(E^u+E^d)(x,0,0)$ and
$x(E^{\bar u}+E^{\bar d})(x,0,0)$. We observe that $x(E^u+E^d)(x,0,0)$
is positive at small $x$, changes sign at $x\approx 0.18$, and is sizeable
and negative at large $x$, taking its maximum around $x\approx 0.5$.
In contrast, $x(E^{\bar u}+E^{\bar d})(x,0,0)$ appears everywhere positive
and rather large at smaller values of $x$. For the second moment of
$(E^u+E^d)(x,0,0)$ we obtain zero (within numerical accuracy) and confirm
numerically the sum rule in Eq.~(\ref{Eq:Eu+Ed-2nd-mom}), which -- let us
stress it -- must be fulfilled in any model which lacks explicit gluon
degrees of freedom but describes the nucleon consistently in terms of
quark- and antiquark degrees of freedom.

In our calculation $x(E^u+E^d)(x,0,0)$ results from the cancellation of
two much larger quantities, namely $xE_M(x,0,0)$ and $x(f_1^u+f_1^d)(x)$,
see Figs.~\ref{Fig:3-spin-mom} and \ref{Fig:4-Eu+Ed-final}, such that one 
could worry whether a small change in parameters could lead to a much 
different final result. The only parameter, which could be in principle 
varied\footnote{
    We work here with the value $M=350\,{\rm MeV}$ which follows from
    instanton phenomenology \cite{Diakonov:1985eg,Diakonov:1983hh}
    and was used in calculations of parton distributions and GPDs
    \cite{Petrov:1998kf,Penttinen:1999th,Schweitzer:2002nm,Diakonov:1996sr,Pobylitsa:1996rs,Diakonov:1997vc,Pobylitsa:1998tk,Weiss:1997rt,Diakonov:1998ze,Wakamatsu:1998rx,Goeke:2000wv}.
    However, in numerous model calculations $M$ was allowed to vary in
    the range $(350-450)\,{\rm MeV}$ \cite{Christov:1995vm}. The value
    $M=420\,{\rm MeV}$ was somehow preferred because it reproduced
    exactly the delta-nucleon mass-splitting within the proper-time 
    regularization \cite{Christov:1995vm}. },
is the constituent quark mass $M$. A calculation with $M=420\,{\rm MeV}$
yields a somehow larger result, see Fig.~\ref{Fig:4-Eu+Ed-final}b. However, 
what is more important in our context, it fully confirms the basic features 
of the calculation with $M=350\,{\rm MeV}$ in Fig.~\ref{Fig:4-Eu+Ed-final}a, 
which is our final result.
Fig.~\ref{Fig:4-Eu+Ed-final}c shows the model results LO-evolved \cite{QCDnum}
to $Q^2=5\,{\rm GeV}^2$ which is a typical scale in experiments.
Hereby we assume the gluon GPD $E^g(x,0,0)$, which mixes with
$(E^u+E^d+E^{\bar u}+E^{\bar d})(x,0,0)$ under evolution, to be zero at the
initial scale taken to be $\mu^2=0.36\,{\rm GeV}^2$. This is justified by the
instanton vacuum model, where twist-2 gluon distributions appear suppressed
with respect to the respective quark distributions. (The meaning of the 
dotted line in Fig.~\ref{Fig:4-Eu+Ed-final}c is explained in the next section.)

  \section{Phenomenological implications}
  \label{Sec:7-asymmetries}

Azimuthal asymmetries in hard exclusive production of vector mesons or pion
pairs from a transversely polarized target are among the observables which
are most sensitive to $E^a(x,\xi,t)$ \cite{Goeke:2001tz}. Let us define
\be\label{Eq:SSA}
    A_{M}(x_B,t,Q^2) = \frac{1}{|S_{\perp}|}
    \frac{\int_{0  }^{ \pi}\di\phi\;\sigma(\phi)
             -\int_{\pi}^{2\pi}\di\phi\;\sigma(\phi)}
         {\int_{0  }^{2\pi}\di\phi\;\sigma(\phi)},
\ee
where $\sigma$ is the cross section for the process
$\gamma_L^\ast(q)+P(p)\to M+P(p^\prime)$ with $\gamma_L^\ast(q)$ denoting
the deeply virtual longitudinally polarized photon with momentum $q$. $P$ is
the incoming (outgoing) proton with momentum $p\,(p^\prime)$, and $M$ denotes
the produced longitudinally polarized vector meson or the pion pair.
For notational simplicity we omit to indicate that $\sigma$ is differential
in $t=(P-P^\prime)^2$, $Q^2=-q^2$, $x_B=Q^2/(2Pq)$ which is related to the
skewedness parameter as $x_B = 2\xi/(1+\xi)$ in the limit $Q^2\to\infty$,
and the angle $\phi$ between the transverse proton spin $S_{\perp}$ and
the plane spanned by the virtual photon and the produced meson.

For not too small $x_B$ the gluon exchange mechanism can be neglected in
a first approximation and the asymmetry $A_M$ reads
\cite{Goeke:2001tz,Lehmann-Dronke:2000xq}
\be\label{eq:asymmetry}
    A_{M}=-\frac{2|\Delta_{\perp}|}{\pi\Mn}\,
    \frac{{\rm Im}(B^{*}C)}{|B|^{2}(1-\xi^{2})-|C|^{2}(\xi^{2}+
    \frac{t}{4M_{N}^{2}})-{\rm Re}(BC^{*})2\xi^{2}}
\ee
where $B$ and $C$ are given for the respective final state $M$ by
\ba
    B_{\rho^{0}} & = &
    \int_{-1}^{1}\di x\;(e_{u}H^{u}-e_{d}H^{d})
    \left(\frac{1}{x-\xi+i\varepsilon}+\frac{1}{x+\xi-i\varepsilon}\right)
    \nonumber \\
    C_{\rho^{0}} & = &
    \int_{-1}^{1}\di x\;(e_{u}E^{u}-e_{d}E^{d})
    \left(\frac{1}{x-\xi+i\varepsilon}+\frac{1}{x+\xi-i\varepsilon}\right)
    \nonumber \\
    B_{\omega} & = &
    \int_{-1}^{1}\di x\;(e_{u}H^{u}+e_{d}H^{d})
    \left(\frac{1}{x-\xi+i\varepsilon}+\frac{1}{x+\xi-i\varepsilon}\right)
    \nonumber \\
    C_{\omega} & = & \int_{-1}^{1}\di x\;(e_{u}E^{u}+e_{d}E^{d})
    \left(\frac{1}{x-\xi+i\varepsilon}+\frac{1}{x+\xi-i\varepsilon}\right)
    \nonumber \\
    B_{\pi^{0}\pi^{0}} & = & \int_{-1}^{1}\di x\;(e_{u}H^{u}+e_{d}H^{d})
    \left(\frac{1}{x-\xi+i\varepsilon}-\frac{1}{x+\xi-i\varepsilon}\right)
    \nonumber \\
    C_{\pi^{0}\pi^{0}} & = & \int_{-1}^{1}\di x\;(e_{u}E^{u}+e_{d}E^{d})
    \left(\frac{1}{x-\xi+i\varepsilon}-\frac{1}{x+\xi-i\varepsilon}\right)
    \,.\label{eq:Amplituden}
\ea
An advantage of choosing such target spin asymmetries to access GPDs
is that the (poorly known) meson distribution amplitudes cancel out.

As we computed here only the forward limit and are particularly interested to
see the impact of our result in comparison to what previously was discussed
in literature, we shall use the method of Ref.~\cite{Radyushkin:1998es} to
generate $\xi$- and $t$-dependence from a given forward limit according to
\be\label{EQ:DD-1}
    E^{q}(x,\xi,t) = E_{DD}^{q}(x,\xi)\;F_2^{q}(t) - \mbox{D-term}
\ee
with the D-term as predicted in the model \cite{Goeke:2001tz},
$F_2(t)$ from \cite{Brash:2001qq,Bosted:1994tm} and
the so-called double distribution given by
\be\label{EQ:DD-2}
    E_{DD}^{q}(x,\xi)=\int_{-1}^{1}\di\beta\int_{-1+|\beta|}^{1-|\beta|}
    \di\alpha\;\delta(x-\beta-\alpha\xi)\,E^q(\beta,0,0)\,h(\beta,\alpha)
    \;.\ee
The Ansatz (\ref{EQ:DD-1},~\ref{EQ:DD-2}) satisfies the polynomiality
condition (\ref{Def:polynom-Eq}) for any ``profile function''
$h(\beta,\alpha)$, e.g., for
\be\label{EQ:DD-3}
    h(\beta,\alpha)=\frac{\Gamma(2b+2)}{2^{2b+1}\Gamma^{2}(b+1)}
    \frac{\left[(1-|\beta|)^{2}-\alpha^{2}\right]^{b}}{(1-|\beta|)^{2b+1}}
    \end{equation}
where we shall choose $b=1$ for our analysis, cf.\  Ref.~\cite{Goeke:2001tz}.
For $(E^u+E^d)(x,0,0)$ we use the result obtained in this work, 
LO-evolved\footnote{
	One could also first construct the GPD at the low scale in the 
        above describe way, and then evolve it to the relevant scale. 
	The difference due to the non-commutativity of these steps is, 
	however, small \cite{Musatov:1999xp} -- far smaller than other 
	uncertainties in our model calculation as well as corrections
	due to the neglect of NLO-corrections, possible power corrections, 
	etc. We shall neglect it here.}
\cite{QCDnum} to $Q^2=5\;{\rm GeV}^2$, cf.\  Fig.~\ref{Fig:4-Eu+Ed-final}c.
For $(E^u-E^d)(x,0,0)$ we shall employ the Ansatz
\be\label{Eq:Eu-Ed}
    (E^u-E^d)(x,0,0)=c_1(f_{1\rm val}^u-f_{1\rm val}^d)(x)+c_2\delta(x)\ee
inspired by the model results for $(E^{u}-E^{d})(x,0,0)$ \cite{Goeke:2001tz}.
In Eq.~(\ref{Eq:Eu-Ed}) $f_{1\rm val}^q$ denotes the respective valence quark
distribution. The parameters $c_{1,2}$ are fixed from
$\kappa^u-\kappa^d=3.706$ and, in order to discuss everything in terms of
results from the model, $J^u-J^d\approx 0.2$ found in the $\chi$QSM 
\cite{Goeke:2001tz}.

\begin{figure}[t!]
\vspace{-0.5cm}
\begin{tabular}{ccc}
    \includegraphics[width=5.4cm,height=4.4cm]{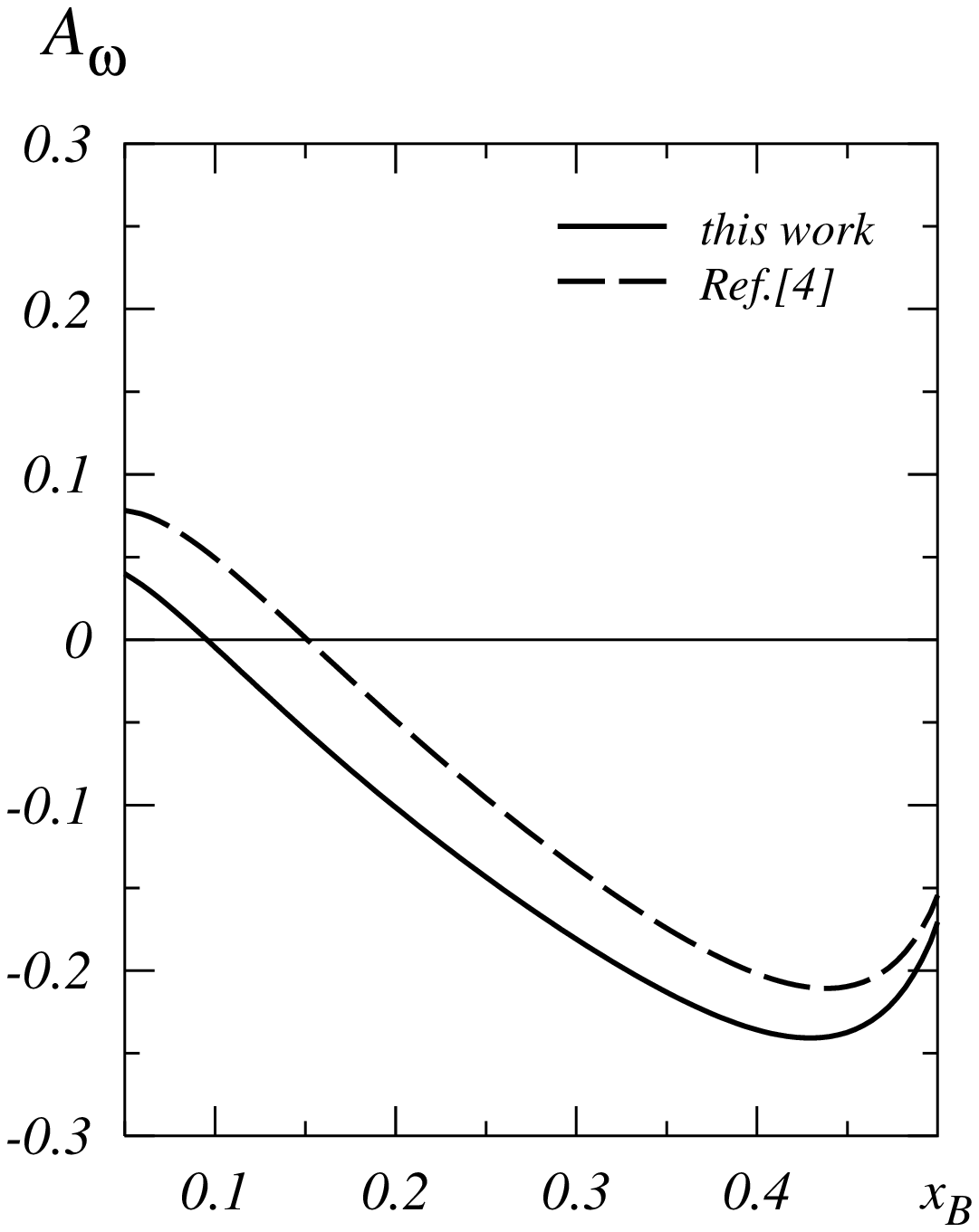}&
    \includegraphics[width=5.4cm,height=4.4cm]{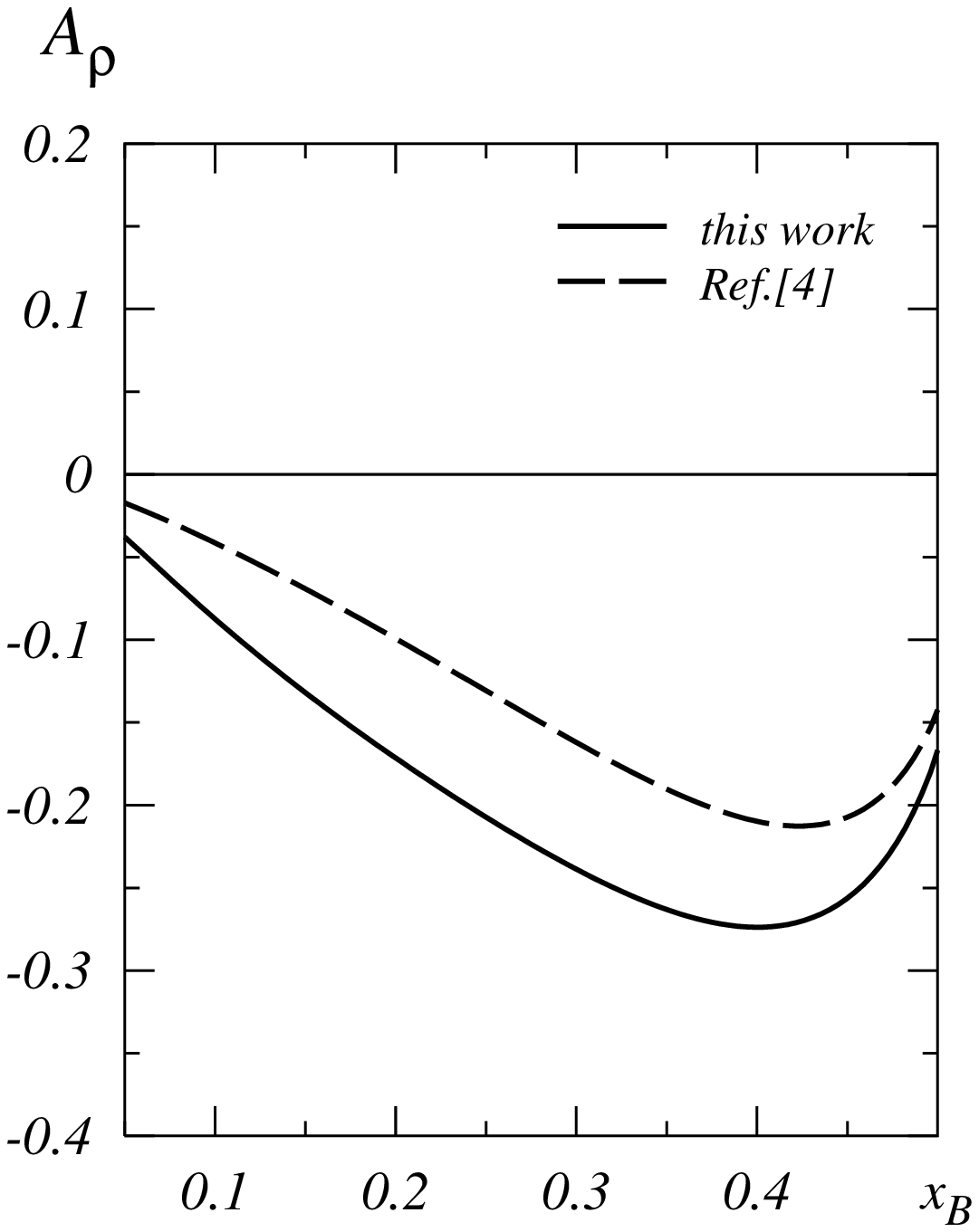}&
    \includegraphics[width=5.4cm,height=4.4cm]{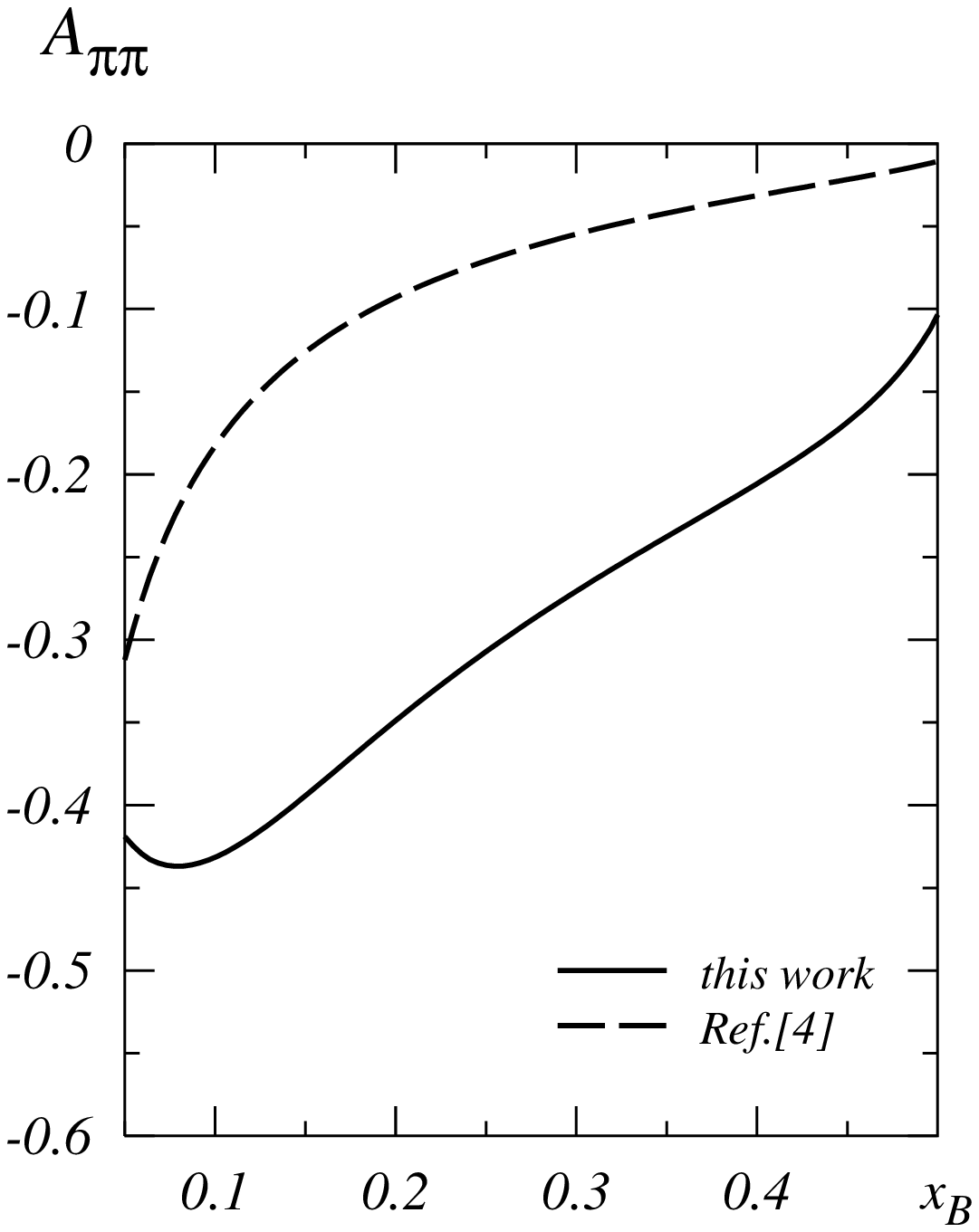}\cr
    \hspace{0.8cm} {\bf a} &
    \hspace{0.8cm} {\bf b} &
    \hspace{0.8cm} {\bf c}
\end{tabular}
    \caption{\footnotesize\label{Fig:5-asymmetries}
    The transverse spin asymmetry in hard exclusive production of
    (a) $\omega^0$, (b) $\rho^0$, and (c) a $\pi^0$ pair as functions
    of $x_B$ at $t=-0.5\;{\rm GeV}^2$ and $Q^2=5\;{\rm GeV}^2$.}
\end{figure}

We shall compare the results obtained with this Ansatz, which is consistently
based on predictions from the $\chi$QSM to the results presented in
Ref.~\cite{Goeke:2001tz} where, in lack of better knowledge, the Ansatz
(\ref{Eq:Eu-Ed}) was assumed to hold also in the flavour singlet case.
Such an assumption is, however, not supported by our direct calculation
as it is demonstrated in Fig.~\ref{Fig:4-Eu+Ed-final}c.

The GPD $H^q(x,\xi,t)$ is modeled in an analog way on the basis of the
parameterization \cite{Martin:2002aw} for $f_1^q(x)$ as described in
Ref.~\cite{Goeke:2001tz}.

As can be seen from Fig.~\ref{Fig:5-asymmetries} the asymmetry has little 
sensitivity to the Ansatz used for $(E^u+E^d)(x,0,0)$ in the case of 
$\omega^0$ or $\rho^0$ production. 
We find that the process most sensitive to $(E^u+E^d)(x,0,0)$ is the 
production of a $\pi^{0}$ pair. The effect of different Ans\"atze can 
cleanly be distinguished.
Experiments at HERMES or JLAB could provide interesting insights. However,
for an unambiguous quantitative analysis of the data it will be necessary
to consider systematically power and NLO corrections.

  \newpage
  \section{Summary and conclusions}
  \label{Sec:8-conclusions}

We have presented in the framework of the flavour-SU(2) version of the
chiral quark-soliton model a study of unpolarized GPDs, which appear at
subleading order in the large-$N_c$ limit, namely the flavour combinations
$(E^u+E^d)(x,\xi,t)$ and $(H^u-H^d)(x,\xi,t)$.
For that we generalized the methods of Ref.~\cite{Petrov:1998kf} developed
to study the  leading large-$N_c$ GPDs, i.e.\  the respectively opposite
flavour combinations.

Interestingly, the GPDs $(H^u+H^d)(x,\xi,t)$ and $(E^u+E^d)(x,\xi,t)$ are of
the same order in the large-$N_c$ counting but handled differently in the
model. The former appears already in leading order, while for the latter
one must invoke $1/N_c$ (rotational) corrections to obtain a non-vanishing
result. Nevertheless they enter the spin sum rule on equal footing,
which is satisfied in the model as we demonstrated. Furthermore, we have
shown that also the subleading GPDs satisfy the polynomiality condition,
and that the coefficients in front of the highest power in $\xi$ in even
moments of $(H^u+H^d)(x,\xi,t)$ and $(E^u+E^d)(x,\xi,t)$ are equal to
each other up to an opposite sign.
Given the different technical handling in the model, the fulfillment of
general requirements where both GPDs are involved is a strong check of
the consistency of the approach.

In the chiral quark-soliton model $M^Q = 2J^Q$ already at a low scale,
i.e.\  the fractions of nucleon momentum and spin carried by quarks and
antiquarks are the same. In QCD this relation becomes exact in the limit
of a large renormalization scale. Actually this is not a specific
prediction of the chiral quark-soliton model, but should hold in any model 
describing the nucleon consistently in terms of quark and antiquark 
degrees of freedom,  i.e.\  without gluon degrees of freedom. 
For such models to be consistent $M^Q=2J^Q=1$ must hold. 
The chiral quark-soliton model fulfills this requirement.
In the framework of the $\chi$QSM the contribution of gluons to the 
nucleon momentum and angular momentum is suppressed with respect to the 
corresponding quark contributions by the small parameter characterizing 
the packing fraction of instantons in vacuum 
\cite{Diakonov:1995qy}.

We then focused on the GPD $(E^u+E^d)(x,\xi,t)$ and computed numerically its
forward limit with a constant (momentum-independent) constituent quark mass
$M$ using the Pauli-Villars regularization. We found that $(E^u+E^d)(x,0,0)$
is negative at larger $x$ and sizeable, and positive at small $x$, changing
the sign at $x\approx 0.2$, while $(E^{\bar u}+E^{\bar d})(x,0,0)$
is always positive and concentrated towards smaller $x$. Noteworthy is that
in the model in the chiral limit $(E^u+E^d)(x,0,0)$ is strongly enhanced
at small $x$ with respect to $(H^u+H^d)(x,0,0)$. It would be interesting
to study in more detail the chiral mechanism responsible for that.

On the basis of our results we estimated azimuthal asymmetries in hard
exclusive production of vector mesons and pion pairs from a transversely
polarized proton target. We observe that pion pair production is very
sensitive to $(E^u+E^d)(x,\xi,t)$. Data from experiments at HERMES or JLAB
on this reaction would be well suited to distinguish among different models.

Our work -- in particular the demonstration of the consistency of the
approach -- not only increases the faith into the applicability of the chiral
quark-soliton model to the computation of GPDs, but also practically paves
the way to further studies of GPDs in leading and subleading order of the
large-$N_c$ limit.

\begin{acknowledgments}
We thank Vadim Guzey for help on evolution, and thank Markus Diehl and Peter
Kroll for valuable discussions. This work is partially supported
by the Graduierten-Kolleg Bochum-Dortmund and Verbundforschung of BMBF.
A Portuguese-German collaboration grant by GRICES and DAAD is acknowledged.
M.~V.~P.\ thanks the Alexander von Humboldt-Stiftung for support by the 
Sofia Kovalevskaya-Award, FNRS (Belgium) and IISN.
\end{acknowledgments}

\subsection*{Note added in proof}

 After this work was finished Ref.~\cite{Diehl:2004cx} appeared, where among 
 others a fit of $(E^q-E^{\bar q})(x,0,t)$ to the Pauli form 
 factor was reported. The result of Ref.~\cite{Diehl:2004cx} for 
 $(E^u+E^d-E^{\bar u}-E^{\bar d})(x,0,0)$ is qualitatively similar to the 
 ansatz of Ref.~\cite{Goeke:2001tz} plotted in Fig.~4c and differs from our 
 result which is negative at all $x$.

\appendix
  \section{Model symmetries}
  \label{App:symmetries}

The effective Hamiltonian $\hat{H}_{\rm eff}$, Eq.~(\ref{Eq:Hamiltonian}),
commutes with the parity operator $\hat{\Pi}$ and the grand-spin operator
\be\label{App:01-grandspin}
    \hat{\bf K}=\hat{\bf S}+\hat{\bf L}+\hat{\bf T}
\ee
where
$\hat{\bf S}=\frac12\gamma_5\gamma^0\bgam$,
$\hat{\bf L}=\hat{\bf r}\times\hat{\bf p}$ and
$\hat{\bf T}=\frac12\btau$ denote the single-quark spin, angular momentum
and isospin operators, respectively.
Thus, the single-quark states are characterized by the eigenvalues of
$\hat{\Pi}$, $\hat{K}^2$ and $K^3$ as $|n\ra=|E_n,\pi,K,K^3\ra$.

Consider the unitary matrix $G_5$ given in the standard representation of
Dirac- and flavour SU(2) matrices by $G_5\equiv \gamma^2\gamma^5\tau^2$.
It acts as $G_5\gamma^\mu G_5^{-1} = (\gamma^\mu)^T$ and
$G_5\tau^a G_5^{-1} = -(\tau^a)^T$ such that in coordinate space the
Hamiltonian (\ref{Eq:Hamiltonian}) is transformed as $G_5 H G_5^{-1}=H^T$.
This transformation acts as $G_5\Phi_n({\bf x})=\Phi_n^\ast({\bf x})$.
The $G_5$ transformation in the effective chiral theory corresponds
to time reversal transformation accompanied by a flavour rotation
\cite{Pobylitsa:2002fr}. Using the $G_5$-transformation one finds
$\la m|\tau^a|j\ra = -\la j|\tau^a|m\ra$ and, e.g.,
\ba
    \la m|(1+\gamma^0\gamma^3)
    F(\hat{\bf p})G(\hat{\bf X})H(\hat{\bf p})|j\ra
 &=&    \la j|(1-\gamma^0\gamma^3)
    H(-\hat{\bf p})G(\hat{\bf X})F(-\hat{\bf p})|m\ra
    \nonumber\\
    \la m|\tau^c(1+\gamma^0\gamma^3)
    F(\hat{\bf p})G(\hat{\bf X})H(\hat{\bf p})|j\ra
 &=&    -\la j|\tau^c(1-\gamma^0\gamma^3)
    H(-\hat{\bf p})G(\hat{\bf X})F(-\hat{\bf p})|m\ra
    \label{App:G5symm}\;.\ea
Another useful symmetry of the model is parity $\hat{\Pi}|m\ra = p_m|m\ra$
with $p_m=\pm 1$ depending on the parity of the state. We obtain, e.g.,
\ba
    \la m|\tau^c(1+\gamma^0\gamma^3)
    F(\hat{\bf p})G(\hat{\bf X})H(\hat{\bf p})|j\ra
 &=&    p_mp_j\la m|\tau^c(1-\gamma^0\gamma^3)
    H(-\hat{\bf p})G(-\hat{\bf X})F(-\hat{\bf p})|j\ra
    \label{App:parity}\;.\ea

  \section{Polynomiality}
  \label{App:polynomiality}

For the discussion of moments the following expressions for
$H_E(x,\xi,t)$ and $E_M(x,\xi,t)$ are more convenient
\ba
 && \hspace{-1cm}
    H_E(x,\xi,t) =
    -\frac{\Mn N_c}{12I}\int\!\frac{\di z^0}{2\pi}\,e^{iz^0x\Mn}\Biggl[
    \biggl\{\doublesum{m,{\rm occ}}{j,{\rm all}}\,e^{-iz^0E_m}
               -\doublesum{m,{\rm all}}{j,{\rm occ}}\,e^{-iz^0E_j}\biggr\}
    \frac{1}{E_m-E_j}\nonumber\\
 &&     \times\,\la m|\tau^a|j\ra\,
        \la j|\,\tau^a(1+\gamma^0\gamma^3)\,\exp(-iz^0\hat{p}^3/2)\,
    \exp(i\bDelta\hat{\bf X})\,\exp(-iz^0\hat{p}^3/2)\,|m\ra
    \nonumber\\
 && +iz^0\singlesum{m,\rm occ}
    e^{-iz^0E_m}\la m|3(1+\gamma^0\gamma^3)\,\exp(-iz^0\hat{p}^3/2)\,
    \exp(i\bDelta\hat{\bf X})\,\exp(-iz^0\hat{p}^3/2)\,|m\ra\Biggr]\;,
    \label{App:EE-model}\\
 && \hspace{-1cm}
    E_M(x,\xi,t) =
        \frac{i\Mn^2N_c}{2I}\int\!\frac{\di z^0}{2\pi}\,e^{iz^0x\Mn}\Biggl[
    \biggl\{\doublesum{m,{\rm occ}}{j,{\rm all}}e^{-iz^0E_m}
               -\doublesum{m,{\rm all}}{j,{\rm occ}}e^{-iz^0E_j}\biggr\}
    \frac{1}{E_m-E_j}\nonumber\\
 &&     \times\,\la m|\tau^b|j\ra\,
        \la j|\,(1+\gamma^0\gamma^3)\,\exp(-iz^0\hat{p}^3/2)\,
        \frac{\epsilon^{3ab} \Delta^a}{\bDelta_\perp^2}\;
    \exp(i\bDelta\hat{\bf X})\,\exp(-iz^0\hat{p}^3/2)\,|m\ra\nonumber\\
 && + iz^0\singlesum{m,\rm occ}e^{-iz^0E_m}
    \la m|\tau^b(1+\gamma^0\gamma^3)\,\exp(-iz^0\hat{p}^3/2)\,
        \frac{\epsilon^{3ab} \Delta^a}{\bDelta_\perp^2}\;
    \exp(i\bDelta\hat{\bf X})\,\exp(-iz^0\hat{p}^3/2)\,|m\ra
    \Biggr]\;.\label{App:EM-model}
\ea
If we take out from the double sums the limiting case $E_m\to E_j$ and combine
it with the respective single sums in (\ref{App:EE-model},~\ref{App:EM-model})
then we arrive at the expressions in Eqs.~(\ref{H_E-model},~\ref{E_M-model}).

\subsection{\boldmath $H_E(x,\xi,t)$}

In order to demonstrate that at $t=0$ the moments of $H_E(x,\xi,t)$ are in
fact functions of $\xi$ only, we have to continue analytically the expression
in Eq.~(\ref{App:EE-model}) to the unphysical point $t=0$.
For that note that
\be
    \exp(i\bDelta\hat{\bf X})
    = \sum\limits_{l_e=0}^\infty
    \frac{(-i2\xi\Mn)^{l_e}}{l_e!}
    |\hat{\bf X}|^{l_e}P_{l_e}(\cos\hat{\theta})
    \label{App:EEmom2}\ee
where $\cos\hat{\theta}=\hat{X}^3/|\hat{\bf X}|$ and $P_{l_e}$ are Legendre
polynomials, {\sl cf.} Ref.~\cite{Schweitzer:2002nm}. Taking moments
of $H_E(x,\xi,t)$ in Eq.~(\ref{App:EE-model}) and using (\ref{App:EEmom2})
we obtain
\ba
&&  M_E^{(L)}(\xi,0)\equiv
    \int\limits_{-1}^1\!\!\di x\;x^{L-1} H_E(x,\xi,t)
    =-\,\frac{\Mn N_c}{12I\Mn^L}
    \sum\limits_{l_e=0}^\infty \frac{(-i2\xi\Mn)^{l_e}}{l_e!}\nonumber\\
&&  \times \Biggl[\sum_{L_i=0}^{L-1}
    \binomial{L-1}{L_i}\sum_{J=0}^{L_i}\binomial{L_i}{J}\frac{1}{2^{L_i}}
    \biggl\{
     \doublesum{m,\rm occ}{j,\rm all}E_m^{L-L_i-1}
        -\doublesum{m,\rm all}{j,\rm occ}E_j^{L-L_i-1}\biggr\}
     \frac{A_{mj}^E}{E_m-E_j}\nonumber\\
&&
    -(L-1)\sum_{L_i=0}^{L-2}\binomial{L-2}{L_i}
    \sum_{J=0}^{L_i}\binomial{L_i}{J}\frac{1}{2^{L_i}}
    \singlesum{m,\rm occ}E_m^{L-L_i-2}
    B^E_{m}\Biggr]
    \;, \label{App:EE-mom3} \ea
with
\ba
&&  A_{mj}^E\equiv
    \la m|\tau^b|j\ra\,\la j|\,\tau^b(1+\gamma^0\gamma^3)\,(\hat{p}^3)^J\,
    |\hat{\bf X}|^{l_e}P_{l_e}(\cos\hat{\theta})
    \,(\hat{p}^3)^{L_i-J}\,|m\ra\;,\nonumber\\
&&  B_m^E\equiv
    \la m|3(1+\gamma^0\gamma^3)\,(\hat{p}^3)^J\,
    |\hat{\bf X}|^{l_e}P_{l_e}(\cos\hat{\theta})
    \,(\hat{p}^3)^{L_i-J}\,|m\ra\;.
    \label{App:EE-mom4}\ea
Using the $G_5$-symmetry, cf.\  Eq.~(\ref{App:G5symm}), we obtain
\ba
&&  A_{mj}^E\equiv (-1)^{L_i}
    \la j|\tau^b|m\ra\,\la m|\,\tau^b(1-\gamma^0\gamma^3)
    \,(\hat{p}^3)^J\,|\hat{\bf X}|^{l_e}P_{l_e}(\cos\hat{\theta})
    \,(\hat{p}^3)^{L_i-J}\,|j\ra\;,\nonumber\\
&&  B_m^E\equiv (-1)^{L_i}
    \la m|3(1-\gamma^0\gamma^3)\,(\hat{p}^3)^J\,
    |\hat{\bf X}|^{l_e}P_{l_e}(\cos\hat{\theta})
    \,(\hat{p}^3)^{L_i-J}\,|m\ra\;.
    \label{App:EE-mom5}\ea
Exploring these identities (for the first it is necessary to rename under the
sum in (\ref{App:EE-mom3}) $m\leftrightarrow j$) we see that $A_{mj}^E$ and
$B_m^E$ in Eq.~(\ref{App:EE-mom3}) are effectively given by (note that
$(\gamma^0\gamma^3)^N=1$ for even $N$ and $\gamma^0\gamma^3$ for odd $N$)
\ba
&&  A_{mj}^E =
    \la m|\tau^b|j\ra\,\la j|\,\tau^b(\gamma^0\gamma^3)^{L_i}
    \,(\hat{p}^3)^J\,|\hat{\bf X}|^{l_e}P_{l_e}(\cos\hat{\theta})
    \,(\hat{p}^3)^{L_i-J}\,|m\ra \;, \nonumber\\
&&  B_m^E\, = \la m|3
    (\gamma^0\gamma^3)^{L_i}\,(\hat{p}^3)^J\,
    |\hat{\bf X}|^{l_e}P_{l_e}(\cos\hat{\theta})
    \,(\hat{p}^3)^{L_i-J}\,|m\ra\;.
    \label{App:EE-mom6}\ea
From parity transformations, cf.\  Eq.~(\ref{App:parity}), we conclude that
$A_{mj}^E=B_m^E=0$ if $l_e$ is odd. Thus we observe that only even powers in
$\xi$ contribute to the moments. In the next step we demonstrate that the
infinite series in $\xi^2$ is actually a polynomial.

For that let us first observe that the object
$\tau^b|j\ra\la j|\tau^b$ (where summation over $b$ is implied) is
invariant under hedgehog rotations. Thus it is practically an irreducible
spherical tensor operator of rank zero with respect to simultaneous isospin-
and space-rotations \cite{Schweitzer:2002nm}.
So in both cases, $A_{mj}^E$ and $B_m^E$, we deal with matrix elements of the
type $\la m|\hat{O}|m\ra$, the only difference being the ``replacement''
$\tau^b|j\ra\la j|\tau^b\rightarrow 3$ which is irrelevant for our arguments.
The matrix elements $\la m|\hat{O}|m\ra$ are non-zero only if the operator
$\hat{O}$ is rank zero, cf.\  \cite{Schweitzer:2002nm} for details.
$\hat{O}$ is a product of $\hat{p}^a$, $\gamma^0\gamma^3$ which have rank 1,
and $P_{l_e}$ which is rank $l_e$, and rank zero operators. $\hat{O}$ is in
general a reducible operator which, however, can be decomposed into a sum of
irreducible tensor operators. For our purposes it is important to find the
highest value for $l_e$ in the Legendre polynomial which yields a
non-vanishing result for $\la m|\hat{O}|m\ra$.
The operators $\hat{p}^a$, which appear $L_i$-times in whatever ordering,
and operator $\gamma^0\gamma^3$, for odd $L_i$, can combine to
an operator with maximally rank $L_i+1$ for odd $L_i$ ($L_i$ for even $L_i$).
Thus $l_e$ can have at most rank $L_i+1$ for odd $L_i$ ($L_i$ for even $L_i$),
otherwise the total operator $\hat{O}$ cannot reach the needed rank zero.
Considering that $L_i\le (L-1)$ we see that the $L^{th}$ moment of
$M_E^{(L)}$ is a polynomial with the highest power $\xi^{L}$ for even $L$
($\xi^{L-1}$ for odd $L$). Thus we obtain
\ba
&&  M_E^{(L)}(\xi,0) = \nonumber\\
&&  -\,\frac{\Mn N_c}{12I\Mn^L} \Biggl[
    \sum_{L_i=0}^{L-1}\binomial{L-1}{L_i}\sum_{J=0}^{L_i}\binomial{L_i}{J}
    \frac{1}{2^{L_i}}\doublesumUp{l_e=0}{l_e\;\rm even}{L_i+1}
    \frac{(-i2\xi\Mn)^{l_e}}{l_e!} 
    \biggl\{\doublesum{m,\rm occ}{j,\rm all}E_m^{L-L_i-1}
           -\doublesum{m,\rm all}{j,\rm occ}E_j^{L-L_i-1}\biggr\}
    \frac{A_{mj}^E}{E_m-E_j}\nonumber\\
&&
    \;\;\;
    -(L-1)\sum_{L_i=0}^{L-2}\binomial{L-2}{L_i}
    \sum_{J=0}^{L_i}\binomial{L_i}{J}\frac{1}{2^{L_i}}
    \doublesumUp{l_e=2}{l_e\;\rm even}{L_i+2}
    \frac{(-i2\xi\Mn)^{l_e-2}}{l_e!}\singlesum{m,\rm occ}E_m^{L-L_i-2}
    B^E_{m}\Biggr]\;. \label{App:EE-mom7} \ea

\subsection{\boldmath $E_M(x,\xi,t)$}

In this case the relevant structure to be continued analytically to $t=0$ is,
{\sl cf.} Ref.~\cite{Schweitzer:2002nm},
\be
    \frac{\Delta^a_\perp}{\bDelta_\perp^2}\;\exp(i\bDelta\hat{\bf X})
    = \sum\limits_{l_e=2}^\infty \frac{(-i2\xi\Mn)^{l_e-2}}{l_e!}
    [\hat{p}^a_\perp, |\hat{\bf X}|^{l_e}P_{l_e}(\cos\hat{\theta})]\;.
        \label{App:EMmom2}\ee
With this result we obtain for the moments of $E_M(x,\xi,t)$ at zero $t$
the result
\ba
&&  M_M^{(L)}(\xi,0)\equiv
    \int\limits_{-1}^1\!\!\di x\;x^{L-1} E_M(x,\xi,t)
    =\frac{i\Mn^2 N_c}{2I\Mn^L}
    \sum\limits_{l_e=2}^\infty \frac{(-i2\xi\Mn)^{l_e-2}}{l_e!}\nonumber\\
&&  \times \Biggl[\sum_{L_i=0}^{L-1}
    \binomial{L-1}{L_i}\sum_{J=0}^{L_i}\binomial{L_i}{J}\frac{1}{2^{L_i}}
    \biggl\{
     \doublesum{m,\rm occ}{j,\rm all}E_m^{L-L_i-1}
        -\doublesum{m,\rm all}{j,\rm occ}E_j^{L-L_i-1}\biggr\}
     \frac{A_{mj}^M}{E_m-E_j}\nonumber\\
&&
    -(L-1)\sum_{L_i=0}^{L-2}\binomial{L-2}{L_i}
    \sum_{J=0}^{L_i}\binomial{L_i}{J}\frac{1}{2^{L_i}}
    \singlesum{m,\rm occ}E_m^{L-L_i-2}
    B_{m}\Biggr]
    \;, \label{App:EM-mom3} \ea
with
\ba
&&  A_{mj}^M\equiv
    \la m|\tau^b|j\ra\,\la j|\,(1+\gamma^0\gamma^3)\,(\hat{p}^3)^J\,
    \epsilon^{3ab}
    [\hat{p}^a_\perp, |\hat{\bf X}|^{l_e}P_{l_e}(\cos\hat{\theta})]
    \,(\hat{p}^3)^{L_i-J}\,|m\ra\;,\nonumber\\
&&  B_m^M\equiv
    \la m|\tau^b\,(1+\gamma^0\gamma^3)\,(\hat{p}^3)^J\,\epsilon^{3ab}
    [\hat{p}^a_\perp, |\hat{\bf X}|^{l_e}P_{l_e}(\cos\hat{\theta})]
    \,(\hat{p}^3)^{L_i-J}\,|m\ra\;.
    \label{App:EM-mom4}\ea
From the $G_5$-symmetry, cf.\  Eq.~(\ref{App:G5symm}), we obtain
\ba
&&  A_{mj}^M =
    \la m|\tau^b|j\ra\,\la j|\,(\gamma^0\gamma^3)^{L_i+1}
    \,(\hat{p}^3)^J\,\epsilon^{3ab}[\hat{p}^a_\perp,
        |\hat{\bf X}|^{l_e}P_{l_e}(\cos\hat{\theta})]
    \,(\hat{p}^3)^{L_i-J}\,|m\ra \;, \nonumber\\
&&  B_m^M\, = \la m|\tau^b\,
    (\gamma^0\gamma^3)^{L_i+1}\,(\hat{p}^3)^J\,\epsilon^{3ab}
    [\hat{p}^a_\perp, |\hat{\bf X}|^{l_e}P_{l_e}(\cos\hat{\theta})]
    \,(\hat{p}^3)^{L_i-J}\,|m\ra\;.
    \label{App:EM-mom6}\ea
From parity transformations, cf.\  Eq.~(\ref{App:parity}), it follows that
$A_{mj}^M$ and $B_m^M$ are zero if $l_e$ is odd. Thus we observe again that
only even powers in $\xi$ contribute to the moments.

In order to demonstrate that the infinite series in even powers of $\xi$
is actually a polynomial, let us consider first the double sum contribution
in Eq.~(\ref{App:EM-mom3}). We deal with the matrix elements $A_{mj}^M$
which are of the type $\la m|\tau^a|j\ra\la j|\hat{O}^a|m\ra$.
The operator $\tau^a$ is an irreducible spherical tensor operator of rank
1 with respect to simultaneous isospin- and space- (hedgehog-) rotations
\cite{Schweitzer:2002nm}. Thus only certain transitions $\la m|\tau^a|j\ra$
between the states $|m\ra$ and $|j\ra$ are non-zero. In order for the
entire expression for $A_{mj}^M$ to be non-zero the same transitions
must appear also in the piece $\la j|\hat{O}^a|m\ra$, thus the operator
$\hat{O}^a$ must be rank 1, too.

The operators $\tau^a$, $\hat{p}^a$, $\gamma^0\gamma^3$ and $P_{l_e}$, which
in our context are relevant for the rank counting, can combine to an operator
with maximally rank $L_i+2$ for even $L_i$ ($L_i+1$ for odd $L_i$).
Thus $l_e$ can be at most $L_i+3$ for even $L_i$ ($L_i+2$ for odd $L_i$),
otherwise the total operator $\hat{O}^a$ cannot reach the needed rank 1.
However, $l_e$ must be even, thus $l_e\le(L_i+2)$ for even $L_i$ ($L_i+1$ for
odd $L_i$). Finally $L_i\le (L-1)$. Therefore the double sum contribution to
the $L^{th}$ moment of $M_M^{(L)}$ is a polynomial with the highest power
$\xi^{L-2}$ for even $L$ ($\xi^{L-1}$ for odd $L$).

$B_m^M$ in Eq.~(\ref{App:EM-mom6}) is zero unless the operators
($|\hat{\bf X}|$, $\tau^a$, the $L_i+1$ operators $\hat{p}^a$, $P_{l_e}$, and
$\gamma^0\gamma^3$ if $L_i$ is odd) combine to a rank 0 operator. One obtains
the same restriction for the maximal value of $l_e$ in term of $L_i$ as above,
however, in this case $L_i\le(L-2)$.  Thus, we finally obtain
(with $A_{mj}^M$, $B_m^M$ given in Eq.~(\ref{App:EM-mom6}))
\ba
&&  M_M^{(L)}(\xi,0) = \nonumber\\
&&  \frac{i\Mn^2 N_c}{2I\Mn^L} \Biggl[
    \sum_{L_i=0}^{L-1}\binomial{L-1}{L_i}\sum_{J=0}^{L_i}\binomial{L_i}{J}
    \frac{1}{2^{L_i}}\doublesumUp{l_e=2}{l_e\;\rm even}{L_i+2}
    \frac{(-i2\xi\Mn)^{l_e-2}}{l_e!}
    \biggl\{\doublesum{m,\rm occ}{j,\rm all}E_m^{L-L_i-1}
           -\doublesum{m,\rm all}{j,\rm occ}E_j^{L-L_i-1}\biggr\}
    \frac{A_{mj}^M}{E_m-E_j}\nonumber\\
&&  \;\;\;
    -(L-1)\sum_{L_i=0}^{L-2}\binomial{L-2}{L_i}
    \sum_{J=0}^{L_i}\binomial{L_i}{J}\frac{1}{2^{L_i}}
    \doublesumUp{l_e=2}{l_e\;\rm even}{L_i+2}
    \frac{(-i2\xi\Mn)^{l_e-2}}{l_e!}\singlesum{m,\rm occ}E_m^{L-L_i-2}
    B_{m}^M\Biggr]\;. \label{App:EM-mom7}
    \ea
We know that $(H^u+H^d)(x,\xi,t)$ satisfies polynomiality and that its
even moments $L$ are polynomials of degree $L$ \cite{Schweitzer:2002nm}.
Here we observe that even moments $L$ of $E_M(x.\xi,t)$ are polynomials of
degree $(L-2)$. Since $E_M(x,\xi,t)$ is the sum of $(E^u+E^d)(x,\xi,t)$ and
$(H^u+H^d)(x,\xi,t)$ this means that the moments of $(E^u+E^d)(x,\xi,t)$
also satisfy polynomiality and that the coefficients of the highest power
$\xi^L$ of even moments satisfy the relation (\ref{Eq:relation-h-e}).

  \section{Proof of the spin sum rule}
  \label{App:proof-of-spin-sum-rule}

The model expression for the second moment of $E_M(x,\xi,t)$ is given by
(\ref{App:EM-mom7}) for $L=2$. By observing that
$[\hat{p}^a_\perp,\hat{\bf X}^2P_2(\cos\hat{\theta})]=i\hat{X}^a_\perp$
we obtain
\be
    M_M^{(2)}(\xi,0)
    = -\,\frac{N_c}{4I}\Biggl[
    \doublesum{m,\rm occ}{j,\rm all}
    \frac{(E_m+E_j)\la m|\tau^b|j\ra\,\la j|\,(\gamma^0\gamma^3)
    \epsilon^{3ab}\hat{X}^a_\perp |m\ra}{E_m-E_j}
    + \doublesum{m,\rm occ}{j,\rm non}
    \frac{2\la m|\tau^b|j\ra\,\la j|\,\hat{p}^3\epsilon^{3ab}
    \hat{X}^a_\perp\,|m\ra }{E_m-E_j}\Biggr]
    \;. \label{App:spin-01}
\ee
Using the following identity, where we explore the equations of motion
in the model,
\be
        \la j|\,(E_m+E_j)\gamma^0\gamma^3\,\hat{X}_\perp^j|m\ra =
    \la j|\,\left\{H\,,\;\gamma^0\gamma^3\,\hat{X}_\perp^j\right\}\,|m\ra=
    \la j|\,\left(2\hat{p}^3\hat{X}_\perp^j+i\gamma^j\gamma^3\right)\,|m\ra
        \label{App:spin-02}
\ee
we obtain
\be
    M_M^{(2)}(\xi,0)
    =\frac{N_c}{2I}\doublesum{m,\rm occ}{j,\rm non}
    \frac{1}{E_m-E_j}\biggl[
      \la m|\tau^3|j\ra\,\la j|\,\gamma^0\gamma^3\gamma_5|m\ra
    +2\la m|\tau^3|j\ra\,\la j|\epsilon^{3ab}\hat{X}^a\hat{p}^b|m\ra\biggr]
    \;.\label{App:spin-03}
\ee
In Eq.~(\ref{App:spin-03}) we identify the model expressions for the spin
$S^Q$ and angular momentum $L^Q$ contribution of quarks and antiquarks to
the spin of the nucleon $S^N$
\ba
&&  S^Q\equiv
    \frac{N_c}{2I}\doublesum{m,\rm occ}{j,\rm non}\frac{1}{E_m-E_j}
    \la m|\tau^3|j\ra\,\la j|\gamma^0\gamma^3\gamma_5|m\ra
    = \frac12\,g_A^{(0)}\;,\nonumber\\
&&  L^Q\equiv
    \frac{N_c}{I}\doublesum{m,\rm occ}{j,\rm non}\frac{1}{E_m-E_j}
    \la m|\tau^3|j\ra\,\la j|\epsilon^{3jk}\hat{X}^j\hat{p}^k|m\ra\;,
    \label{App:spin-04}
\ea
where $g_A^{(0)}$ denotes the isosinglet axial coupling constant.
The operators in (\ref{App:spin-03}) can be added according to
Eq.~(\ref{App:01-grandspin})
\be
    \frac12\gamma^0\gamma^3\gamma_5+\epsilon^{3jk}\hat{X}^j\hat{p}^3
    =\hat{S}^3+\hat{L}^3=\hat{K}^3-\hat{T}^3\;.
\ee
Noting that in Eq.~(\ref{App:spin-03}) the matrix elements
$\la j|\hat{K}^3|m\ra = K^3_j\delta_{jm}$ vanish we obtain
\be
    M_M^{(2)}(\xi,0) \equiv 2S^Q+2L^Q
    =\frac{N_c}{4I}\doublesum{m,\rm occ}{j,\rm non}
    \frac{1}{E_j-E_m}\la m|\tau^3|j\ra\,\la j|\tau^3|m\ra = 1
    \;.\label{App:spin-05}
\ee


\end{document}